\documentclass[aps,preprint,amsmath,amssymb,showpacs,amsfonts,prd,nofootinbib]{revtex4-1}
\usepackage{epsfig}
\usepackage{graphicx}
\usepackage{dcolumn}
\usepackage{bm}
\usepackage{amsmath, amsthm, amssymb}
\usepackage{color}

\numberwithin{equation}{section}

\newcommand{\rank}{\operatorname{rank}}
\newcommand{\del}{\partial}
\newcommand{\Lie}[1]{\mathcal{L}_{#1}\,}
\newcommand{\Liebold}[1]{\Lie{\mathbf{#1}}}
\newcommand{\dd}[3]{\left(\frac{\del #1}{\del #2}\right)_{\! #3}} 
\newtheorem{proposition}{Proposition}[section]

\newcommand{\calH}{\mathcal{H}}
\newcommand{\calT}{\mathcal{T}}
\newcommand{\calG}{\mathcal{G}}
\newcommand{\Jcs}{J_{\mathrm{CS}}}
\newcommand{\dds}[1]{\dd{#1}{s}{\mathbf{n}/s}}
\newcommand{\ddn}[2]{\dd{#1}{n^{#2}}{s}}

\begin{document}
\title{Holographic Non-Abelian Charged Hydrodynamics\\ from the Dynamics of Null Horizons}

\author{Christopher Eling$^1$}
\author{Yasha Neiman$^2$}
\author{Yaron Oz$^2$}
\affiliation{$^1$ SISSA, Via Bonomea 265, 34136 Trieste, Italy and INFN Sezione di Trieste, Via Valerio 2, 34127 Trieste, Italy}
\affiliation{$^2$ Raymond and Beverly Sackler School of
Physics and Astronomy, Tel-Aviv University, Tel-Aviv 69978, Israel}

\date{\today}

\begin{abstract}

We analyze the dynamics of a four-dimensional null hypersurface in a five-dimensional  bulk spacetime with Einstein-Yang-Mills fields. In an appropriate ansatz, the projection of the field equations onto the hypersurface takes the form of conservation laws for relativistic hydrodynamics with global non-abelian charges. A Chern-Simons term in the bulk action corresponds to anomalies in the global charges, with a vorticity  term arising in the hydrodynamics. We derive
the entropy current and obtain unique expressions for some of the leading-order transport coefficients (in the abelian case, all of them) for arbitrary equations of state. As a special case and a concrete example, we discuss the event horizon of a boosted Einstein-Yang-Mills black brane in an asymptotically Anti-de-Sitter spacetime. The evolution equations in that case describe the hydrodynamic limit of a conformal field theory with anomalous global non-abelian charges on the Anti-de-Sitter boundary.

\end{abstract}
\pacs{04.70.-s, 47.10.ad, 11.25.Tq }
\maketitle

\section{Introduction and Summary} \label{sec:intro}

The hydrodynamics of relativistic conformal field theories (CFTs) has attracted much attention, largely in view of the AdS/CFT correspondence between gravitational theories on asymptotically Anti-de-Sitter (AdS) spaces and CFTs \cite{Maldacena:1997re} (for a review see \cite{Aharony:1999ti}). Hydrodynamics is an effective description of the long distance field theory dynamics and applies under the condition that the correlation length of the fluid $l_{cor}$ is much smaller than the characteristic scale $L$ of variations of the macroscopic fields. The AdS/CFT correspondence implies that the long wavelength
dynamics of gravity is dual to the CFT hydrodynamics. Indeed, it has been shown in \cite{Bhattacharyya:2008jc} that the $(d+1)$-dimensional CFT hydrodynamics equations are the same as the equations describing the evolution of large scale perturbations of the $(d+2)$-dimensional black brane. The derivation in \cite{Bhattacharyya:2008jc} is similar to
the derivation of hydrodynamics from the Boltzmann equation, where the
thermal equilibrium solution is the boosted black brane \cite{Fouxon:2008ik}. The limit of non-relativistic macroscopic motions in CFT hydrodynamics leads to the non-relativistic incompressible
Navier-Stokes (NS) equations \cite{Fouxon:2008tb,Bhattacharyya:2008kq}, and the dual gravitational description is found by taking the non-relativistic limit of the geometry dual to the relativistic CFT hydrodynamics \cite{Bhattacharyya:2008kq}.

In \cite{Eling:2009pb} the membrane paradigm formalism \cite{Damour,Price:1986yy,membrane} and an expansion in powers of the Knudsen number $l_{cor}/L$ have been used to show that the dynamics of a membrane defined by the event horizon of a black brane in asymptotically AdS space-time is described by the incompressible Navier-Stokes equations of non-relativistic fluids. This has been generalized in \cite{Eling:2009sj} to the relativistic CFT hydrodynamics case. The starting point in the membrane paradigm framework is an equilibrium $(d+2)$-dimensional solution containing a timelike Killing vector field and a stationary $(d+1)$-dimensional causal horizon. This solution is associated with a thermal state at uniform temperature. When a hydrodynamic limit exists, we can expand the general inhomogeneous black brane in the neighborhood of the causal horizon in powers of $l_{cor}/L$, assuming that there is no singularity at the horizon. For a  black brane in asymptotically AdS, the set of Einstein equations projected into the horizon surface at lowest orders in $l_{cor}/L$ is equivalent to the $(d+1)$-dimensional relativistic CFT Navier-Stokes equations.
The membrane paradigm has previously been used as a tool for calculating transport coefficients \cite{Kovtun:2003wp, Saremi:2007dn, Fujita:2007fg, Starinets:2008fb, Iqbal:2008by, Paulos:2009yk}.
The RG flow relation between the horizon coefficients and those of the boundary gauge theory have been recently discussed in \cite{Bredberg:2010ky}.

It has been recently revealed that the hydrodynamics description exhibits an interesting effect when a global abelian symmetry current of the microscopic theory is anomalous. This has been first discovered in the context of the fluid/gravity correspondence \cite{Erdmenger:2008rm,Banerjee:2008th}. The Chern-Simons term in the gravity action, which corresponds to having an anomalous global abelian symmetry current in the dual gauge theory, has been shown to modify the hydrodynamic current by a term proportional to the vorticity of the fluid. At first sight the additional vorticity term seemed in contradiction with the second law of thermodynamics \cite{Landau}. This, however, has been resolved by a redefinition of the entropy current in \cite{Son:2009tf}.
Experimental signals of the vorticity term from heavy ion collisions have been proposed in \cite{KerenZur:2010zw,Kharzeev:2010gr}.

 In this paper will analyze the dynamics of a four-dimensional null hypersurface in a five-dimensional  bulk spacetime with Einstein-Yang-Mills fields. In an appropriate ansatz, the projection of the field equations onto the hypersurface takes the form of conservation laws for relativistic hydrodynamics with global non-abelian charges. A Chern-Simons term in the bulk action corresponds to anomalies in the global charges, with a vorticity  term arising in the hydrodynamics. We will derive
the entropy current and obtain unique expressions for some of the leading-order transport coefficients (in the abelian case, all of them) for arbitrary equations of state. As a special case and a concrete example, we will consider  the event horizon of a boosted Einstein-Yang-Mills black brane in an asymptotically Anti-de-Sitter spacetime. The evolution equations in that case describe the hydrodynamic limit of a conformal field theory with anomalous global non-abelian charges on the Anti-de-Sitter boundary.

We work perturbatively in the derivative/Knudsen number expansion, and write the equations to second order; equivalently, we consider the hydrodynamic currents and stress tensor to the first viscous order. We place all 4d fields on the null hypersurface, without explicit reference to the bulk spacetime or its boundary.
When an AdS asymptotics is used for the unperturbed homogeneous brane solution, it provides the equation of state for the thermal system.
Without solving the radial equations, we consider the most general inhomogeneous corrections to the horizon projections of the bulk fields. These corrections are then constrained by the requirement that the 4d constraint equations take the form of conservation laws. In the abelian case, the constitutive relations are obtained from this procedure in a unique closed form. For non-abelian charges, some of the terms are similarly obtained in closed form, with part of the conductivity matrix remaining ambiguous.

The null surface formalism introduced here improves on the methods developed in \cite{Eling:2009sj} by utilizing an intrinsic covariant divergence on the horizon. Our geometrical calculation is performed with the horizon's null normal as the basic vector variable. In viscous hydrodynamics, this corresponds to the ``frame" where the entropy velocity is the basic variable; however, the results are readily translated into the more standard Landau frame, which uses the energy velocity.

The horizon dynamics equations define the viscous hydrodynamics of a relativistic CFT with non-abelian conserved currents and generalizes the known cases with conserved abelian currents. In the presence of a non-abelian Chern-Simons term, the horizon dynamics gives the hydrodynamics equations in the presence of anomalous non-abelian global symmetries. The result generalizes the hydrodynamics equations of field theories with triangle anomalies of abelian currents, discussed above.

The thermodynamics of the AdS black brane is described by the equation of state
\begin{align}
	\epsilon(s,n^a) = \frac{3}{\pi}\left(\frac{s}{2}\right)^{4/3}\left(1 + \frac{\pi^2 n_a n^a}{3s^2}\right), \label{eq:state_intro}
\end{align}
where $\epsilon$, $s$ and $n^a$ are the energy, entropy and charge densities, respectively. The temperature $T$, chemical potentials $\mu_a$ and pressure $p$ are derived from the equation of state through the usual thermodynamic identities.

In the present calculation, we will make no use of the particular equation of state \eqref{eq:state_intro}. Instead, we study a generalized ansatz for the horizon fields, corresponding to an arbitrary equation of state. Thus, in addition to the concrete case of the AdS black brane, we study more generally the structure of hydrodynamic systems which can be encoded holographically on a null hypersurface with Einstein-Yang-Mills fields and an Einstein-Yang-Mills-Chern-Simons action. In particular, do not use the conformal symmetry of the AdS asymptotics and of eq. \eqref{eq:state_intro}.

We now proceed to a summary of the results, expressed in the Landau frame. For the viscous stress-energy density, we find the familiar expression
\begin{align}
	T^{\mu\nu} = \sqrt{-h}\left(p(h^{\mu\nu} + 4u^\mu u^\nu) - \frac{s}{2\pi}\pi^{\mu\nu}\right),
\label{eq:T_intro}
\end{align}
where $u^\mu$ is the energy 4-velocity, and $h_{\mu\nu}$ is the 4d background metric for the hydrodynamics. Our formalism is such that a curved $h_{\mu\nu}$ can be allowed for with no additional complications. $\pi_{\mu\nu}$ is the shear tensor, defined as
\begin{align}
	\pi_{\mu\nu} = P^\rho_\mu P^\sigma_\nu D_{(\rho}u_{\sigma)} - \frac{1}{3}P_{\mu\nu} D_\rho u^\rho,
\end{align}
where $P_{\mu\nu} \equiv h_{\mu\nu} + u_{\mu}u_{\nu}$ is the projector orthogonal to $u^\mu$, and $D_\mu$ is the covariant derivative with respect to $h_{\mu\nu}$. Eq. \eqref{eq:T_intro} implies the celebrated value $\eta = \frac{s}{4\pi}$ for the shear viscosity, and a vanishing bulk viscosity $\zeta = 0$.

For the (non-abelian) charge currents, we obtain the expression
\begin{align}
  \begin{split}
  J_a^\mu ={}& \sqrt{-h}\left(n_a u^\mu
    - T\left(\frac{s^{1/3}}{2^{4/3}\pi}\left(Z_{ab} - \frac{2\mu_{(a} n_{b)}}{\epsilon + p} + \frac{\mu_c\mu^c n_a n_b}{(\epsilon + p)^2}\right)
      + \sigma^\bot_{ab}\right)P^{\mu\nu} D_\nu\frac{\mu^b}{T}\right) \\
    &- \frac{2}{\pi}\left(\beta_{abc}\mu^b\mu^c - \frac{2n_a\beta_{bcd}\mu^b\mu^c\mu^d}{3(\epsilon + p)}\right)\omega^\mu, \label{eq:J_intro}
  \end{split}
\end{align}
where $\omega^\mu = \frac{1}{2}\epsilon^{\mu\nu\rho\sigma}u_\nu\del_\rho u_\sigma$ is the fluid's vorticity, $\beta_{abc}$ are the Chern-Simons/anomaly coefficients, and $Z_{ab}$ is the projector onto the charge subalgebra $Z(\mathbf{n})$ which commutes with the local charge density $n^a$. The distinctively non-abelian contribution is concentrated in the coefficient matrix $\sigma^\bot_{ab}$, whose indices lie in the orthogonal complement $Z^\bot(\mathbf{n})$ of $Z(\mathbf{n})$. Its symmetric part $\sigma^\bot_{(ab)}$ is positive semi-definite, as required by the second law of thermodynamics. Other than that, our method places no restriction on the function $\sigma^\bot_{ab}(s,n^c)$. The constitutive relation \eqref{eq:J_intro} reduces to the known results \cite{Erdmenger:2008rm,Banerjee:2008th} in the abelian case, where we substitute $Z_{ab} = \delta_{ab}$ and $\sigma^\bot_{ab} = 0$. The abelian case is also reproduced in the limit of small charges or weak couplings, as we will see in section \ref{sec:discussion}.

We will find that the undetermined part $\sigma^\bot_{ab}$ of the conductivity matrix depends on the inhomogeneous corrections to the bulk gauge potential. In the AdS/CFT context, these are fixed by the radial part of the field equations together with the condition:
\begin{align}
  \lim_{r\rightarrow\infty}A^a_\mu = 0\ . \label{eq:A_infty}
\end{align}
On the event horizon of the AdS black brane, the resulting corrections to $A^a_\mu$ are given by some functional of the hydrodynamic fields. Without solving the radial equations, we cannot know the details of this functional. Outside the AdS context, our result is simply that the transport coefficient $\sigma^\bot_{ab}$ can be arbitrary in Einstein-Yang-Mills-Chern-Simons hydrodynamic systems (and at leading order, it is the \emph{only} arbitrary transport coefficient). In the AdS/CFT case, we obtain algebraic relations between the corrections to the non-abelian gauge potential $A^a_\mu$ on the horizon and the conserved current $J^\mu_a$. This current is given on the other hand by the AdS/CFT prescription as:
\begin{align}
	J_\mu^a = -\frac{1}{2\pi}\lim_{r\rightarrow\infty}r^2 A_\mu^a\ . \label{eq:J_boundary}
\end{align}
Thus, we obtain algebraic relations between the corrections to the gauge potential on the horizon and on the boundary. We verified these relations numerically
against the bulk first-order calculation for the $SU(2)$ case carried out in \cite{Torabian:2009qk}.

Finally, our result for the entropy current reads:
\begin{align}
  s^\mu = \sqrt{-h}\left(su^\mu + \frac{s^{4/3}T}{2^{4/3}\pi(\epsilon + p)}\left(\mu^a - \frac{\mu_b\mu^b n^a}{\epsilon + p}\right) P^{\mu\nu}D_\nu\frac{\mu_a}{T}\right)
    + \frac{4s\beta_{abc}\mu^a\mu^b\mu^c}{3\pi(\epsilon + p)}\omega^\mu\ . \label{eq:s_intro}
\end{align}
This result generalizes the entropy current derived previously on the gravity side of AdS/CFT using properties of the uncharged black brane solution \cite{Bhattacharyya:2008xc}.

The focusing equation of the null horizon requires the divergence of this entropy current to be non-negative, which translates into an entropy production rate:
\begin{align}
  \del_\mu s^\mu = \frac{1}{4}\del_\mu S^\mu
    = \sqrt{-h}\left(T\sigma_{ab} P^{\mu\nu} D_\mu\frac{\mu^a}{T}D_\nu\frac{\mu^b}{T} + \frac{s}{2\pi T}\pi_{\mu\nu}\pi^{\mu\nu}\right).
\end{align}

The Chern-Simons/anomaly terms in \eqref{eq:J_intro} and \eqref{eq:s_intro} reproduce the results of the general thermodynamic argument in \cite{Son:2009tf}, which was in turn motivated by the specific results from the AdS black brane. As mentioned above, our calculation yields these results for an arbitrary equation of state. Furthermore, we generalize the work in \cite{Son:2009tf} by considering hydrodynamics with non-abelian as well as abelian charges.

The paper is organized as follows.  In section \ref{sec:framework} we present the geometrical framework, we define the fields and field equations in the 5d bulk and their projections onto the event horizon.  Section \ref{sec:homogeneous} describes the homogeneous (thermodynamic) black brane solution from which we begin and its various parameters, and outlines our approach to the inhomogeneous brane. In section \ref{sec:ideal}, we derive the leading-order (ideal) hydrodynamic equations. The principal weight of the paper lies in section \ref{sec:viscous}, where the second-order (viscous) hydrodynamic equations are derived. In section \ref{sec:discussion}, we discuss the results. In appendix \ref{sec:notation}, we provide a summary of the notation used in the paper.

\section{The Geometrical Framework} \label{sec:framework}

We consider the Einstein-Yang-Mills fields: the metric $g_{AB}$ and the gauge potential $A^a_A$ for a gauge group $\calG$. These fields live in a 5d spacetime with a null horizon $\calH$. We consider the relevant restrictions of the fields into the 4d geometry of the horizon, and develop the general formalism which will be applied to the hydrodynamic ansatz in the following sections.

\subsection{Horizon geometry} \label{sec:framework:field_eqs:geometry}

In the bulk spacetime, we choose coordinates of the form $x^A = (r, x^\mu)$. The $x^\mu$ coordinatize the horizon $\calH$; $r$ is a transverse coordinate, with $r = 0$ on $\calH$. $\del_A r$ is a null covector tangent to the $\calH$. When raised with the metric, it gives a vector field $\ell^A = g^{AB}\del_B r$ which is both normal and tangent to $\calH$, and tangent to its null generators. In components, we have $\ell^A = (0, \ell^\mu)$. These choices fix the following components of the inverse bulk metric on $\calH$:
\begin{align}
	g^{rr} = 0; \quad g^{r\mu} = \ell^\mu\ . \label{eq:g^rA}
\end{align}

The pullback of $g_{AB}$ into $\calH$ is the degenerate horizon metric $\gamma_{\mu\nu}$. Its null directions are the generating light-rays of $\calH$, i.e. $\gamma_{\mu\nu}\ell^\nu = 0$. The Lie derivative of $\gamma_{\mu\nu}$ along $\ell^\mu$ gives us the shear/expansion tensor, or ``second fundamental form'':
\begin{align}
	\theta_{\mu\nu} \equiv \frac{1}{2}\Lie{\ell}\gamma_{\mu\nu}\ , \label{eq:theta_def}
\end{align}
which has the properties:
\begin{align}
	\theta_{\mu\nu} = \theta_{\nu\mu}; \quad \ell^\mu\theta_{\mu\nu} = 0\ . \label{eq:theta_props}
\end{align}
We can write a decomposition of $\theta_{\mu\nu}$ into a shear tensor $\sigma^{(H)}_{\mu\nu}$ and an expansion coefficient $\theta$:
\begin{align}
	\theta_{\mu\nu} = \sigma^{(H)}_{\mu\nu} + \frac{1}{3}\theta\gamma_{\mu\nu}\ . \label{eq:theta_decompose}
\end{align}

An equivalent condition which we found useful in practice is
\begin{align}
	\theta = \lambda(G^{-1})^{\mu\nu}\theta_{\mu\nu}\ , \label{eq:theta_trace}
\end{align}
where $(G^{-1})^{\mu\nu}$ is the inverse of any matrix $G_{\mu\nu}$ of the form:
\begin{align}
	G_{\mu\nu} = \lambda\gamma_{\mu\nu} - b_\mu b_\nu; \quad b_\mu\ell^\mu \neq 0\ . \label{eq:G}
\end{align}
We introduced the superfluous scalar field $\lambda$ for later convenience: it will turn out that a matrix of the form \eqref{eq:G} coincides at leading order with the metric $h_{\mu\nu}$ of the hydrodynamic dual.

The expansion coefficient $\theta$ is related to the horizon area density current $S^\mu$, which is the analog of the bulk volume density $\sqrt{-g}$. This current is a vector density tangent to the null generators, with magnitude defined by:
\begin{align}
	 \epsilon^{\mu_0\mu_1\mu_2\mu_3}\gamma_{\mu_1\nu_1}\gamma_{\mu_2\nu_2}\gamma_{\mu_3\nu_3} = S^{\mu_0}S^{\nu_0}\epsilon_{\nu_0\nu_1\nu_2\nu_3}\ .
\end{align}
Here, $\epsilon^{\mu\nu\rho\sigma}$ is the invariant 4d Levi-Civita density with components $\pm 1$, and $\epsilon_{\mu\nu\rho\sigma}$ is the inverse density with components $\pm 1$. We have the collinearity relation $S^\mu = v\ell^\mu$, where $v$ ($v\equiv 4 s$) is a scalar density. Note that unlike $\ell^\mu$, the magnitude of $S^\mu$ is fixed uniquely by the horizon metric.
 In our adapted bulk coordinates, the value of the 4d density $v(x^\mu)$ equals that of the 5d density $\sqrt{-g}$ evaluated at the corresponding horizon point $(0,x^\mu)$. Finally, the divergence of the area current $S^\mu$ is related to $\theta$ by:
\begin{align}
	v\theta = \del_\mu S^\mu\ . \label{eq:theta_S}
\end{align}

Since $\gamma_{\mu\nu}$ is degenerate, one cannot use it to define an intrinsic connection on the null horizon, as could be done for spacelike or timelike hypersurfaces. The bulk spacetime's connection does induce a notion of parallel transport in $\calH$, but only along its null generators. This structure is not fully captured by $\gamma_{\mu\nu}$; instead, it is encoded by the extrinsic curvature, or 'Weingarten map' $\Theta_\mu{}^\nu$, which is the horizon restriction of $\nabla_A\ell^B$:
\begin{align}
	\Theta_\mu{}^\nu = \nabla_\mu\ell^\nu\ .
\end{align}
For a non-null hypersurface, the extrinsic curvature at a point is independent of the induced metric. For null hypersurfaces, this is not so. Indeed, lowering an index on $\Theta_\mu{}^\nu$ with $\gamma_{\mu\nu}$ (and losing information in the process), we get the shear/expansion tensor $\theta_{\mu\nu}$:
\begin{align}
	\Theta_\mu{}^\rho\gamma_{\rho\nu} = \theta_{\mu\nu}\ . \label{eq:Theta_theta}
\end{align}
This expresses the compatibility of the parallel transport defined by $\Theta_\mu{}^\nu$ with the horizon metric $\gamma_{\mu\nu}$. Contracting $\Theta_\mu{}^\nu$ with $\ell^\mu$ yields the surface gravity $\kappa$, which measures the non-affinity of $\ell^\mu$:
\begin{align}
	\Theta_\mu{}^\nu\ell^\mu = \kappa\ell^\nu\ . \label{eq:Theta_kappa}
\end{align}
It follows from \eqref{eq:Theta_theta}-\eqref{eq:Theta_kappa} that given an arbitrary $G_{\mu\nu}$ of the form \eqref{eq:G}, $\Theta_\mu{}^\nu$ can be written as:
\begin{align}
	\Theta_\mu{}^\nu = \lambda\theta_{\mu\rho}(G^{-1})^{\rho\nu} + c_\mu\ell^\nu; \quad c_\mu\ell^\mu = \kappa\ . \label{eq:Theta}
\end{align}
The covector $c_\mu$ encodes the 4 degrees of freedom in $\Theta_\mu{}^\nu$ which are independent of $\gamma_{\mu\nu}$. In the hydrodynamics, these degrees of freedom roughly correspond to the velocity and temperature fields.

The parallel transport along null generators defined by $\Theta_\mu{}^\nu$ can possess a Ricci-type curvature, described by a 4d covector density. In bulk terms, this quantity is given by the projection $R_{\mu\nu}S^\nu$ of the bulk Ricci tensor $R_{AB}$; therefore, it is coupled to matter by the Einstein equation. On the other hand, it is related to $\Theta_\mu{}^\nu$ through a null version of the Gauss-Codazzi equations. We will now describe this relation, and bring it to a convenient form. Our discussion follows and expands on the formalism of \cite{Jezierski:2001qv}.

In their usual form, the Gauss-Codazzi equations involve the divergence of the hypersurface's extrinsic curvature. Our null horizon, however, does not possess an intrinsic connection, and a covariant divergence of $\Theta_\mu{}^\nu$ cannot be defined. The solution is to define a tensor \emph{density} constructed out of $\Theta_\mu{}^\nu$ in the following manner:
\begin{align}
	Q_\mu{}^\nu = v(\Theta_\mu{}^\nu - \kappa\delta_\mu^\nu),
\end{align}
with the properties:
\begin{align}
	\ell^\mu Q_\mu{}^\nu = 0; \quad Q_\mu{}^\rho\gamma_{\rho\nu} = Q_\nu{}^\rho\gamma_{\rho\mu}\ . \label{eq:Q_props}
\end{align}
It turns out that the degenerate metric $\gamma_{\mu\nu}$ is sufficient to define the divergence of any density which satisfies \eqref{eq:Q_props}. We begin by defining a raised-index version of $Q_\mu{}^\nu$, which satisfies:
\begin{align}
	 Q^{\mu\nu} =  Q^{\nu\mu}; \quad \gamma_{\mu\rho} Q^{\rho\nu} = Q_\mu{}^\nu\ . \label{eq:upper_Q}
\end{align}
The properties \eqref{eq:upper_Q} do not define $ Q^{\mu\nu}$ uniquely, but only up to multiples of $\ell^\mu\ell^\nu$; this will be good enough for our purpose. The horizon-intrinsic covariant divergence of $Q_\mu{}^\nu$ can now be defined as follows:
\begin{align}
	\bar{\nabla}_\nu Q_\mu{}^\nu = \del_\nu Q_\mu{}^\nu - \frac{1}{2} Q^{\nu\rho}\del_\mu\gamma_{\nu\rho}\ . \label{eq:div_Q}
\end{align}
Since the contraction of $\del_\mu\gamma_{\nu\rho}$ with $\ell^\nu\ell^\rho$ is zero\footnote{This follows directly from the identity $\del_\mu(\gamma_{\nu\rho} \ell^\nu \ell^\rho) = 0$}, the potential ambiguity in its contraction with $Q^{\nu\rho}$ drops out, and we have a well-defined expression. Using this divergence, the null Gauss-Codazzi equation can be written as:
\begin{align}
	R_{\mu\nu}S^\nu = \bar{\nabla}_\nu Q_\mu{}^\nu - v\del_\mu\theta\ . \label{eq:GC_stub}
\end{align}

We found it convenient to use the decomposition \eqref{eq:Theta} of $\Theta_\mu{}^\nu$ when evaluating eq. \eqref{eq:GC_stub}. The corresponding decomposition of $Q_\mu{}^\nu$ reads:
\begin{align}
	Q_\mu{}^\nu = \lambda v\theta_{\mu\rho}(G^{-1})^{\rho\nu} + (c_\mu S^\nu - c_\rho S^\rho \delta_\mu^\nu), \label{eq:Q_decomposition}
\end{align}
where each of the two terms separately satisfies eqs. \eqref{eq:Q_props}. We can therefore consider their covariant divergences separately. Let's start with the first term. A raised-index version of $\lambda v\theta_{\mu\rho}(G^{-1})^{\rho\nu}$ satisfying \eqref{eq:upper_Q} can be obtained by simply raising with $\lambda(G^{-1})^{\mu\nu}$; furthermore, eq. \eqref{eq:theta_props} and the relation $(G^{-1})^{\mu\nu}b_\nu \sim \ell^\mu$ allow us to replace $\del_\mu\gamma_{\nu\rho}$ in \eqref{eq:div_Q} with $\del_\mu(\lambda^{-1}G_{\nu\rho})$. We get:
\begin{align}
	\bar\nabla_\nu\left(\lambda v\theta_{\mu\rho}(G^{-1})^{\rho\nu}\right) = D^{(G)}_\nu\left(\lambda v\theta_{\mu\rho}(G^{-1})^{\rho\nu}\right) + v\theta\del_\mu\ln\sqrt{\lambda},
\end{align}
where $D^{(G)}_\mu$ is the covariant derivative associated with the ``metric'' $G_{\mu\nu}$. The divergence of the second term in \eqref{eq:Q_decomposition} is a simpler matter; an object of this type has a covariant divergence which depends only on $S^\mu$ and $c_\mu$, with no need for the full $\gamma_{\mu\nu}$:
\begin{align}
	\bar\nabla_\nu(c_\mu S^\nu - c_\rho S^\rho \delta_\mu^\nu) = c_\mu \del_\nu S^\nu + 2S^\nu\del_{[\nu}c_{\mu]}\ . \label{eq:div_c_independent}
\end{align}
Summing up, we have:
\begin{align}
	R_{\mu\nu}S^\nu = D^{(G)}_\nu\left(\lambda v\theta_{\mu\rho}(G^{-1})^{\rho\nu}\right) + v\theta\del_\mu\ln\sqrt{\lambda}
		+ c_\mu \del_\nu S^\nu + 2S^\nu\del_{[\nu}c_{\mu]} - v\del_\mu\theta\ . \label{eq:GC_basic}
\end{align}
The $S^\mu$-component of $R_{\mu\nu}S^\nu$ can be expressed more directly, as the null focusing equation:
\begin{align}
	R_{\mu\nu}\ell^\mu\ell^\nu = -\ell^\mu\del_\mu\theta + \kappa\theta - \frac{1}{3}\theta^2
		- \lambda^2(G^{-1})^{\mu\rho}(G^{-1})^{\nu\sigma}\sigma^{(H)}_{\mu\nu}\sigma^{(H)}_{\rho\sigma}\ . \label{eq:focusing_basic}
\end{align}
Once again, the shear-squared term could be defined more directly using the tangent bundle modulo displacements along light-rays. The form given here is more convenient for practical calculations.

\subsection{Yang-Mills} \label{sec:framework:field_eqs:ym}

The bulk gauge potential $A^a_A$ is associated with a field strength $F^a_{AB} = 2\del_{[A}A^a_{B]} + f^a{}_{bc}A^b_A A^c_B$. The structure constants $f_{abc} = f_{[abc]}$ include the coupling strengths of each simple piece of the gauge group. The gauge group in the 5d spacetime induces a 4d gauge group on $\calH$. The corresponding gauge potential $A^a_\mu$ is the pullback of the bulk potential $A^a_A$ into $\calH$. Likewise, the field strength $F^a_{\mu\nu} = 2\del_{[\mu}A^a_{\nu]} + f^a{}_{bc}A^b_\mu A^c_\nu$ associated with $A^a_\mu$ is the pullback of $F^a_{AB}$.

Another restriction of $F^a_{AB}$ into $\calH$ with a well-defined 4d meaning is given by $S^\nu F_{a\nu}{}^A$. This quantity is tangent to the horizon. The upper index can therefore be restricted into the tangent bundle of $\calH$, giving a 4d vector density. In our adapted coordinates, it equals $\sqrt{-g}F_a^{r\mu}$. Inserting a numerical factor, we define:
\begin{align}
	j^{a\mu} \equiv \frac{\sqrt{-g}}{4\pi}F^{ar\mu} = \frac{v}{4\pi}\ell^\nu(F^a_{\nu r}\ell^\mu + F^a_{\nu\rho}g^{\rho\mu}), \label{eq:j}
\end{align}
where we used \eqref{eq:g^rA} in the second equality. At the ideal-fluid approximation, this will coincide with the hydrodynamic charge current. The numerical factor is chosen to make the chemical potential of the fluid agree with the ``electric potential'' $\ell^\mu A^a_\mu$ on $\calH$. Note that $j_a^\mu$ is covariant under 4d gauge transformations on $\calH$. The two quantities $A^a_\mu$ and $j_a^\mu$ on the horizon play a role similar to that of the non-normalizable and normalizable parts of $A^a_\mu$ on the AdS boundary. The quantities $F^a_{\mu\nu}$ and $j_a^\mu$ are not independent, and do not span all the components of $F^a_{AB}$ on $\calH$. This is a consequence of the null nature of $\calH$. Explicitly, we have the following relation, analogous to \eqref{eq:Theta_theta}:
\begin{align}
  \gamma_{\mu\nu}j^{a\nu} = \frac{1}{4\pi}S^\nu F^a_{\nu\mu}\ . \label{eq:j_F}
\end{align}
The single degree of freedom in $j_a^\mu$ which is independent of $F^a_{\mu\nu}$ corresponds to the hydrodynamic charge density.

\subsection{Field equations} \label{sec:framework:field_eqs:equations}

\subsubsection{Bulk equations}

The bulk fields satisfy the Einstein equation with a cosmological constant and a Yang-Mills stress tensor:
\begin{align}
	& R_{AB} - \Lambda g_{AB} = 8\pi\calT_{AB}\ , \label{eq:Einstein} \\
	& \calT_{AB} \equiv \frac{1}{4\pi}(F^a_{AC} F_{aB}{}^C - \frac{1}{6} F^a_{CD} F_a^{CD} g_{AB}),
\end{align}
coupled to the Yang-Mills equation with a Chern-Simons term:
\begin{align}
	\del_B\left(\sqrt{-g}F_a^{AB}\right) + \sqrt{-g}f_{ab}{}^c A^b_B F_c^{AB} - \beta_{abc}\epsilon^{ABCDE} F^b_{BC} F^c_{DE} = 0\ , \label{eq:YM-CS}
\end{align}
where $\epsilon^{ABCDE}$ is the invariant 5d Levi-Civita density with components $\pm 1$. The Chern-Simons coefficients $\beta_{abc} = \beta_{(abc)}$ can be any gauge-invariant constant tensor. In particular, they can mix different $U(1)$ and simple pieces of the gauge group $\calG$. Examples include:
\begin{itemize}
	\item Any constant tensor for Abelian $\calG$.
	\item $\beta_{abc} \sim d_{abc}$ for $\calG = SU(N)$, $N \geq 3$.
	\item $\beta_{abc} \sim \delta^0_{(a}\delta^{d'}_b\delta^{d'}_{c)}$ for gauge group $\calG = \calG' \times U(1)$ with $\calG'$ simple, where $0$ is the index of the $U(1)$ generator, and $d'$ runs over the generators of $\calG'$.
\end{itemize}
In general, the gauge-invariance of $\beta_{abc}$ is encoded by the condition:
\begin{align}
	f^{ed}{}_{a}\beta_{dbc} + f^{ed}{}_{b}\beta_{adc} + f^{ed}{}_{c}\beta_{abd} = 0\ . \label{eq:beta_invariance}
\end{align}

\subsubsection{Horizon equations} \label{sec:framework:field_eqs:equations:horizon}

To restrict the Einstein equation \eqref{eq:Einstein} into $\calH$, we contract with $S^\nu$ and pull back the remaining index. The result is an equation between 4d covector densities. The $\sim g_{AB}$ terms don't contribute, since $g_{A\nu}S^\nu \sim g_{AB}\ell^B = \del_A r$, and that vanishes under pullback. We get:
\begin{align}
	R_{\mu\nu}S^\nu = 8\pi\calT_{\mu\nu}S^\nu\ . \label{eq:Einstein_contracted}
\end{align}
The LHS can be expressed in terms of lower-derivative geometric quantities on $\calH$ via the Gauss-Codazzi equation \eqref{eq:GC_basic}. For the RHS, we have, using \eqref{eq:g^rA} and \eqref{eq:j}:
\begin{align}
	\calT_{\mu\nu}S^\nu = \frac{v}{4\pi}F^a_{\mu C}F_{a\nu}{}^C\ell^\nu = F^a_{\mu\nu}j_a^\nu\ . \label{eq:bulk_T_S}
\end{align}
Substituting into \eqref{eq:Einstein_contracted}, we get the Einstein-Gauss-Codazzi equation:
\begin{align}
	D^{(G)}_\nu\left(\lambda v\theta_{\mu\rho}(G^{-1})^{\rho\nu}\right) + v\theta\del_\mu\ln\sqrt{\lambda} + c_\mu \del_\nu S^\nu + 2S^\nu\del_{[\nu}c_{\mu]} - v\del_\mu\theta
		= 8\pi F^a_{\mu\nu}j_a^\nu\ . \label{eq:GC}
\end{align}
The contraction of this equation with $\ell^\mu/v$ can be written more simply using the focusing equation \eqref{eq:focusing_basic}. We get:
\begin{align}
	-\ell^\mu\del_\mu\theta + \kappa\theta - \frac{1}{3}\theta^2 - \lambda^2(G^{-1})^{\mu\rho}(G^{-1})^{\nu\sigma}\sigma^{(H)}_{\mu\nu}\sigma^{(H)}_{\rho\sigma} 	 
		= \frac{8\pi}{v}\ell^\mu F^a_{\mu\nu} j_a^\nu\ . \label{eq:focusing}
\end{align}
The Yang-Mills-Chern-Simons equation \eqref{eq:YM-CS} can be restricted to $\calH$ by contraction with $\epsilon_{A\mu\nu\rho\sigma}\epsilon^{\mu\nu\rho\sigma}/24$. Dividing further by $4\pi$, we write the resulting equation as:
\begin{align}
  \del_\mu(j^{a\mu} + \Jcs^{a\mu}) + f^a{}_{bc} A^b_\mu(j^{c\mu} + \Jcs^{c\mu}) + \frac{\beta^a{}_{bc}}{2\pi}\epsilon^{\mu\nu\rho\sigma}f^b{}_{de}A^d_\mu A^e_\nu F^c_{\rho\sigma}
    = 0\ , \label{eq:YM_horizon}
\end{align}
where $j^{a\mu}$ is defined in \eqref{eq:j}, and $\Jcs^{a\mu}$ is a Chern-Simons current:
\begin{align}
	\Jcs^{a\mu} = -\frac{\beta^a{}_{bc}}{2\pi}\epsilon^{\mu\nu\rho\sigma} A^b_\nu F^c_{\rho\sigma}\ . \label{eq:j_CS}
\end{align}
To derive the Chern-Simons part of \eqref{eq:YM_horizon}, we used the identity $\del_{[\mu}F^a_{\nu\rho]} + f^a{}_{bc}A^b_{[\mu}F^c_{\nu\rho]} = 0$ and eq. \eqref{eq:beta_invariance}. At the leading viscous order, $\Jcs^{a\mu}$ will play the role of the vorticity term in the hydrodynamic charge current.

\section{The AdS black brane} \label{sec:homogeneous}

\subsection{The homogeneous solution}

For Anti-de-Sitter asymptotics, we choose a negative cosmological constant $\Lambda = -4$ in \eqref{eq:Einstein}. The field equations \eqref{eq:Einstein}-\eqref{eq:YM-CS} then admit a family of exact solutions, describing a charged black brane in equilibrium. The solutions are parameterized by a set of constants: a 4d symmetric form $h_{\mu\nu}$ with Lorentzian signature, a velocity $\ell^\mu$, a mass parameter $m$ and a charge parameter $q^a$ in the adjoint representation of $\calG$. In (radially shifted) Eddington-Finkelstein coordinates, the solution reads:
\begin{align}
	g_{AB}dx^A dx^B &= -2\ell_\mu dx^\mu dr + (R + r)^2\left(P_{\mu\nu} - V(R + r, m, q^a)\ell_\mu\ell_\nu\right) dx^\mu dx^\nu\ , \label{eq:BBmetric} \\
	A^a_A dx^A &= -\frac{\sqrt{3}q^a}{2(R + r)^2} \ell_\mu dx^\mu\ . \label{eq:Agauge}
\end{align}
The 4d indices $(\mu,\nu,\dots)$ are implicitly lowered and raised with the boundary metric $h_{\mu\nu}$ and its inverse $h^{\mu\nu}$. We may consider $h_{\mu\nu}$ as a flat non-degenerate metric over the $x^\mu$. The velocity $\ell^\mu$ is normalized with respect to this metric, so that $h_{\mu\nu}\ell^\mu\ell^\nu = \ell_\mu\ell^\mu = -1$. $P_{\mu\nu} = h_{\mu\nu} + \ell_\mu\ell_\nu$ is the projector orthogonal to $\ell^\mu$ with respect to $h_{\mu\nu}$. The function $V(\bar{r}, m, q^a)$ is given by:
\begin{align}
	V(\bar{r},m,q) = 1 - \frac{m}{\bar{r}^4} + \frac{q_a q^a}{\bar{r}^6}\ .
\end{align}
$\bar{r} = R(m,q^a)$ denotes the largest root of $V(\bar{r},m,q)$, i.e.:
\begin{align}
	m = R^4\left(1 + \frac{q_a q^a}{R^6}\right)\ . \label{eq:m}
\end{align}
This relation makes $(R, q^a)$ more convenient than $(m, q^a)$ as basic variables. We will find that these variables are related to the fluid's entropy and charge densities.

The solution \eqref{eq:BBmetric}-\eqref{eq:Agauge} is homogeneous in all but the radial direction. It does not depend on the gauge structure constants  and couplings, or the Chern-Simons coefficients. The hypersurface $\calH$ defined by $r = 0$ is the brane's outer event horizon. In accord with the notation of section \ref{sec:framework:field_eqs:geometry}, the vector $\ell^A = (0, \ell^\mu)$ is normal to the horizon, and we have $g_{AB}\ell^B = \del_A r$ on $\calH$.

Substitution and direct calculation give us the horizon quantities defined in sections \ref{sec:framework:field_eqs:geometry}-\ref{sec:framework:field_eqs:ym}:
\begin{align}
  \gamma^{(0)}_{\mu\nu} &= R^2 P_{\mu\nu}; & S^{(0)\mu} &= \sqrt{-h}R^3\ell^\mu; & v^{(0)} &= \sqrt{-h}R^3; \\
  \Theta_\mu^{(0)\nu} &= -\kappa^{(0)}\ell_\mu\ell^\nu; & \kappa^{(0)} &= R\left(2 - \frac{q_a q^a}{R^6}\right); \\
  A^{(0)a}_\mu &= -\frac{\sqrt{3}q^a}{2R^2}\ell_\mu; & j_a^{(0)\mu} &= \frac{\sqrt{3}}{4\pi}\sqrt{-h}q_a\ell^\mu\ .
\end{align}
Here and below, all quantities are evaluated on the horizon unless otherwise specified. The '(0)' superscripts anticipate the inhomogeneous corrections. To calculate $\Theta_\mu^{(0)\nu}$ and $j_a^{(0)\mu}$, we used the horizon values of the derivatives $\del_r g_{\mu\nu}$ and $\del_r A^a_\mu$, respectively. These are the only $r$-derivatives we will use throughout the paper; the calculation of inhomogeneous corrections will focus on $\calH$ exclusively. In the homogeneous solution \eqref{eq:BBmetric}-\eqref{eq:Agauge}, all partial derivatives with respect to $x^\mu$ vanish; also, we have $f_{abc}A^{(0)b}_{[\mu}A^{(0)c}_{\nu]} = 0$. Therefore:
\begin{align}
	\theta^{(0)} = 0; \quad \sigma^{(H)(0)}_{\mu\nu} = 0; \quad F^{(0)a}_{\mu\nu} = 0; \quad \Jcs^{(0)a\mu} = 0\ .
\end{align}

We'd also like to have a convenient matrix of the form \eqref{eq:G}. Let us choose $\lambda = 1/R^2$ and $b_\mu = \ell_\mu$, so that:
\begin{align}
	G_{\mu\nu} \equiv \frac{1}{R^2}\gamma_{\mu\nu} - \ell_\mu\ell_\nu\ .
\end{align}
For the solution \eqref{eq:BBmetric}-\eqref{eq:Agauge}, this simply equals $h_{\mu\nu}$:
\begin{align}
	G^{(0)}_{\mu\nu} = h_{\mu\nu}\ . \label{eq:G_0}
\end{align}
In the decomposition \eqref{eq:Theta} of $\Theta_\mu{}^\nu$, we have only the $c_\mu$-term, with the value:
\begin{align}
	c^{(0)}_\mu = -\kappa^{(0)}\ell_\mu\ .
\end{align}

\subsection{Thermodynamic quantities} \label{sec:homogeneous:thermo}

Our black brane in equilibrium can be viewed as a thermodynamic system living in the 4d metric $h_{\mu\nu}$. The velocity vector $\ell^\mu$ defines the system's rest frame. Using the Bekenstein-Hawking correspondence between area and entropy, we define an entropy current as:
\begin{align}
	s^\mu = \frac{1}{4}S^\mu = \sqrt{-h}s\ell^\mu\ ,
\end{align}
where the entropy density $s$ is given by:
\begin{align}
	s = \frac{v^{(0)}}{4\sqrt{-h}} = \frac{1}{4}R^3\ . \label{eq:s_0}
\end{align}
The temperature is given by the surface gravity:
\begin{align}
	T = \frac{\kappa^{(0)}}{2\pi} = \frac{R}{2\pi}\left(2 - \frac{q_a q^a}{R^6}\right). \label{eq:T}
\end{align}
The energy density can now be derived as:
\begin{align}
	\epsilon = \int_0^{s}{Tds'} = \frac{3R^4}{16\pi}\left(1 + \frac{q_a q^a}{R^6}\right) = \frac{3m}{16\pi}\ . \label{eq:epsilon} 
\end{align}

In equilibrium, we identify $j_a^\mu$ with the system's charge current $J_a^\mu$. This will be justified by a conservation law in section \ref{sec:ideal:current}. The charge density is then given by:
\begin{align}
	n^a = \frac{\sqrt{3}q^a}{4\pi}\ , \label{eq:n}
\end{align}
so that:
\begin{align}
	J^{(0)\mu}_a = j^{(0)\mu}_a = \sqrt{-h}n_a\ell^\mu\ . \label{eq:J_0_n}
\end{align}
The chemical potential and pressure can then be derived as:
\begin{align}
	\mu_a &= \left(\frac{\del\epsilon}{\del n^a}\right)_s = \frac{\sqrt{3}q_a}{2R^2} = \left.\ell^\mu A^{(0)}_{a\mu}\right|_{\calH}\ , \\
	p &= Ts + \mu_a n^a - \epsilon = \frac{R^4}{16\pi}\left(1 + \frac{q_a q^a}{R^6}\right) = \frac{m}{16\pi}\ . \label{eq:p}	
\end{align}

Finally, we define the system's stress-energy density as:
\begin{align}
	T_\mu^{(0)\nu} = \sqrt{-h}(p P_\mu^\nu + \epsilon\ell_\mu\ell^\nu) = \sqrt{-h}\left(p\delta_\mu^\nu + (\epsilon + p)\ell_\mu\ell^\nu\right). \label{eq:T_0}
\end{align}
Once the parameters $(R,q^a)$ are eliminated, our knowledge about the brane's thermodynamics is fully described by the equation of state:
\begin{align}
	\epsilon(s,n^a) = \frac{3}{\pi}\left(\frac{s}{2}\right)^{4/3}\left(1 + \frac{\pi^2 n_a n^a}{3s^2}\right), \label{eq:state}
\end{align}
which can be derived from \eqref{eq:s_0}, \eqref{eq:epsilon} and \eqref{eq:n}. This equation of state is conformally covariant, which is also evident from the tracelessness relation $\epsilon = 3p$.

Using the thermodynamic variables, the horizon quantities from the previous subsection can be rewritten as:
\begin{align}
	\gamma^{(0)}_{\mu\nu} &= (4s)^{2/3} P_{\mu\nu}; & c^{(0)}_\mu &= -2\pi T\ell_\mu; \label{eq:grav_0} \\
	A^{(0)a}_\mu &= -\mu^a\ell_\mu; & j_a^{(0)\mu} &= \sqrt{-h}n_a\ell^\mu\ , \label{eq:charge_0}
\end{align}
where $c_\mu$ is again defined using the auxiliary matrix:
\begin{align}
  G_{\mu\nu} = \frac{1}{(4s)^{2/3}}\gamma_{\mu\nu} - \ell_\mu\ell_\nu \qquad (\lambda = \frac{1}{(4s)^{2/3}};\ b_\mu = \ell_\mu)\ , \label{eq:G_specific}
\end{align}
with $G^{(0)}_{\mu\nu} = h_{\mu\nu}$. The other quantities can be derived directly from \eqref{eq:grav_0}:
\begin{align}
  S^{(0)\mu} &= 4\sqrt{-h}s\ell^\mu; & v^{(0)} &= 4\sqrt{-h}s; \label{eq:S_0} \\
  \Theta_\mu^{(0)\nu} &= -2\pi T\ell_\mu\ell^\nu; & \kappa^{(0)} &= 2\pi T\ . \label{eq:Theta_0}
\end{align}
In the following, we will use the ansatz \eqref{eq:grav_0}-\eqref{eq:charge_0} directly, with no reference to the specific equation of state \eqref{eq:state}, or to conformal invariance. We only demand $T$ and $\mu_a$ to be the derivatives of \emph{some} energy function $\epsilon(s,n^a)$ with respect to $s$ and $n^a$, respectively.

\subsection{Inhomogeneous solution: outline of the calculation}

Consider again the configuration \eqref{eq:BBmetric}-\eqref{eq:Agauge}, but with $(h_{\mu\nu},\ell^\mu,R,q^a)$ slowly varying functions of $x^\mu$ rather than constants. We will interpret this $x^\mu$-dependence by treating $(h_{\mu\nu},\ell^\mu,R,q^a)$ (or $(h_{\mu\nu},\ell^\mu,s,n^a)$) as fields on the brane's horizon $\calH$. Various quantities and equations can be expanded order by order in powers of the small $\del_\mu$ derivatives (note, however, that $\del_r$ derivatives are not small). We use the symbolic small parameter $\varepsilon$ to count these powers. We will refer to the power of $\varepsilon$ involved as the ``order'' of a quantity or an equation.

In general, \eqref{eq:BBmetric}-\eqref{eq:Agauge} with $x^\mu$-dependent parameters will not be a solution of the field equations. However, for certain 4d configurations of $(h_{\mu\nu},\ell^\mu,s,n^a)$, it will approximate a solution. This has been shown for trivial, $U(1)$ and $SU(2)$ gauge groups, and we assume this to be the case for general gauge groups as well. The horizon fields of the exact solution are given by functionals of $(h_{\mu\nu},\ell^\mu,s,n^a)$, for which eqs. \eqref{eq:grav_0}-\eqref{eq:charge_0} provide the zeroth-order terms.

The constraints on the 4d fields $(h_{\mu\nu},\ell^\mu,s,n^a)$ arise from the horizon projections \eqref{eq:GC}, \eqref{eq:YM_horizon} of the Einstein and Yang-Mills equations. We will begin by writing the first-order constraints for the ansatz \eqref{eq:grav_0}-\eqref{eq:charge_0} (ideal equations), followed by the second-order constraints using the most general corrections to \eqref{eq:grav_0}-\eqref{eq:charge_0} (viscous equations). We require the constraint equations to take the form of conservation laws with respect to the metric $h_{\mu\nu}$. For the ideal equations, this requirement is automatically satisfied; however, at the viscous order, it restricts the possible corrections to the ansatz \eqref{eq:grav_0}-\eqref{eq:charge_0}. The equations then describe the dynamics of a 4d fluid living in the metric $h_{\mu\nu}$, with configuration $(\ell^\mu,s,n^a)$. The null normal $\ell^\mu$ corresponds to the fluid's entropy velocity (since $S^\mu \sim \ell^\mu$ is the area current). Most of the remaining freedom in the corrections to \eqref{eq:grav_0}-\eqref{eq:charge_0} has no effect on the hydrodynamic transport coefficients. The corrections to $A^a_\mu$ are the exception, since they affect the non-abelian parts of the conductivity. Thus, we obtain a substantial part of the transport coefficients in a unique closed form.

Our analysis will take place on the event horizon $\calH$ of the exact inhomogeneous solution. In the configuration \eqref{eq:BBmetric}-\eqref{eq:Agauge} with inhomogeneous $(h_{\mu\nu},\ell^\mu,s,n^a)$, the hypersurface $r=0$ is null, and doesn't intersect the singularity. It is therefore an event horizon for the zeroth-order solution, and a zeroth-order approximation for the horizon of the corresponding exact solution. For the corrected solution, we will still use coordinates so that $r = 0$ and $g^{rA} = \ell^A$ on $\calH$.

Alternatively, our calculation can be viewed from a more general perspective, as an analysis of the evolution equations of a null hypersurface in a hydrodynamic ansatz. The equations for the AdS black brane are then obtained as a special case.

\section{Ideal hydrodynamics} \label{sec:ideal}

\subsection{Charge tensors constructed from $n^a$} \label{sec:ideal:q_tensors}

The hydrodynamic charge density $n^a$ defines a distinguished generator in the gauge algebra. Before we proceed with the ideal hydrodynamic equations, we pause to analyze the simplest charge tensors that can be built from $n^a$ in a $\calG$-covariant manner. These are given by various products of $n^a$ with the gauge group's constant invariant tensors, such as $f_{abc}$ and $\delta_{ab}$. A reader interested only in the abelian case may skip to section \ref{sec:ideal:current}.

\subsubsection{Scalars}

There are $\rank\calG$ functionally independent charge scalars which can be constructed from $n^a$. They are obtained by saturating the indices of totally-symmetric constant invariant $\calG$-tensors with factors of $n^a$. We denote a full set of such independent scalars by $N_i$, where the index $i$ runs from $1$ to $\rank\calG$. The values of $N_i$ determine $n^a$ up to gauge rotations. In the abelian case, the components of $n^a$ are themselves a full set of invariant scalars.

\subsubsection{Vectors}

There are $\rank\calG$ linearly independent charge vectors which can be constructed from $n^a$. As a basis for these vectors, we may use $(N_i)^a = \del N_i/\del n_a$. Because they are constructed from $n^a$ and constant invariant tensors, the $(N_i)^a$ all commute with $n^a$ and with each other under the gauge algebra. For non-singular values of $n^a$, the $(N_i)^a$ form a basis for the centralizer $Z(\mathbf{n})$ of $n^a$, i.e. the subspace of the gauge algebra which commutes with $n^a$.

\subsubsection{Matrices} \label{sec:homogeneous:q_tensors:matrices}

One type of charge matrix which can be built out of $n^a$ is the external product $(N_i)^a (N_j)^b$ of two vectors. For generic values of $n^a$, such matrices span the entire matrix space over $Z(\mathbf{n})$. In particular, their linear combination gives the projector $Z_{ab}$ onto $Z(\mathbf{n})$ with respect to the Killing metric (for an abelian gauge group we define $Z_{ab} = \delta_{ab}$, and similarly for abelian pieces of a semi-simple group).

There are also charge matrices which do not lie in $Z(\mathbf{n})$, and are not reducible into products of covariantly constructed vectors. To clarify the structure of such matrices $M_{ab}$, note that for any vector $(N_i)^b$, the expression $M_{ab}(N_i)^b$ is a charge vector constructed from $n^a$. Therefore, $M_{ab}(N_i)^b$ lies in $Z(\mathbf{n})$. It follows that for generic values of $n^a$, the matrix $M_{ab}$ can be expressed in block-diagonal form:
\begin{align}
	M_{ab} = M^{||}_{ab} + M^\bot_{ab}\ ,
\end{align}
where both indices of $M^{||}_{ab}$ and $M^\bot_{ab}$ lie respectively in $Z(\mathbf{n})$ and $Z^\bot(\mathbf{n})$, the orthogonal complement of $Z(\mathbf{n})$. The projector $Z^\bot_{ab} = \delta_{ab} - Z_{ab}$ onto $Z^\bot(\mathbf{n})$ is an example of a $M^\bot_{ab}$-type matrix. Another trivial example is $f_{abc}n^c$.

The matrix $M_{ab}$ is constructed only from $n^a$ and invariant constant tensors; therefore, it must be invariant under rotations generated by $n^a$. The corresponding infinitesimal statement is:
\begin{align}
	f_{adc}n^c M_{db} + f_{bdc}n^c M_{ad} = 0\ \Rightarrow\ f_{adc}n^c M_{db} - M_{ad} f_{dbc}n^c = 0\ . \label{eq:M_commute}
\end{align}
In other words, $M_{ab}$ commutes with $f_{abc}n^c$ in the matrix sense. For the $M^{||}_{ab}$-piece, this is trivial. For the $M^\bot_{ab}$-piece, the commutation implies that it must be a matrix polynomial in $f_{abc}n^c$. This is because for generic $n^a$, the matrix $f_{abc}n^c$ has distinct eigenvalues over $Z^\bot(\mathbf{n})$, since are no multiplicities in the nontrivial roots of compact simple Lie groups. But if any $M^\bot_{ab}$-type matrix is a polynomial in $f_{abc}n^c$, then these matrices all commute with each other. Since they also commute with all $M^{||}_{ab}$-type matrices, we have:
\begin{proposition} \label{prop:commutation}
  All matrices constructed from $n^a$ with indices in $Z^\bot(\mathbf{n})$ commute with all other matrices constructed from $n^a$
\end{proposition}

\subsection{Current conservation} \label{sec:ideal:current}

At first order, the restricted Yang-Mills equation \eqref{eq:YM_horizon} can be written as:
\begin{align}
	\del_\mu j_a^{(0)\mu} + f_{abc}(A^{(0)b}_\mu j^{(1)c\mu} - j^{(0)b\mu} A^{(1)c}_\mu) = O(\varepsilon^2), \label{eq:div_J_0_raw}
\end{align}
where ``(1)'' superscripts indicate first-order corrections to the zeroth-order expressions \eqref{eq:charge_0}.

\subsubsection{Abelian case}

For abelian $\calG$, the $f_{abc}$-terms in \eqref{eq:div_J_0_raw} vanish, and we immediately get the ideal current conservation equation:
\begin{align}
	\del_\mu j_a^{(0)\mu} = O(\varepsilon^2). \label{eq:div_J_0}
\end{align}
We can now substitute the partial derivative with $D_\mu$, the covariant derivative with respect to $h_{\mu\nu}$. Plugging in the zeroth-order value from section \eqref{sec:homogeneous:thermo} and dividing by $\sqrt{-h}$, the equation becomes:
\begin{align}
	Dn_a + n_a D_\mu\ell^\mu = O(\varepsilon^2), \label{eq:D_q_0}
\end{align}
where $D \equiv \ell^\mu D_\mu$ is the directional derivative along $\ell^\mu$, and $D_\mu\ell^\mu = P^{\mu\nu}D_\mu\ell_\nu$ is the fluid's expansion rate. Eq. \eqref{eq:div_J_0} justifies our interpretation \eqref{eq:J_0_n} of $j_a^{(0)\mu}$ as the zeroth-order conserved charge current $J^{(0)\mu}_a$.

\subsubsection{Non-abelian case}

For a non-abelian charge group, the same ideal conservation law is obtained. First, let us expand the $f_{abc}$-terms in \eqref{eq:div_J_0_raw} as:
\begin{align}
  f_{abc}(A^{(0)b}_\mu j^{(1)c\mu} - j^{(0)b\mu} A^{(1)c}_\mu) = -f_{abc}(\mu^b\ell_\mu j^{(1)c\mu} + n^b\ell^\mu A^{(1)c}_\mu). \label{eq:f_terms_1}
\end{align}
Let us contract this with a charge vector $(N_i)^a$ which commutes with $n^a$. The second term in \eqref{eq:f_terms_1} vanishes immediately. As for the first term, note that $\mu_a$ is a $\calG$-covariant function of $n^a$ and the scalar $s$. It is therefore in the centralizer $Z(\mathbf{n})$ of $n^a$, and commutes with $(N_i)^a$. We conclude that contraction with $(N_i)^a$ annihilates both terms in \eqref{eq:f_terms_1}. Since the vectors $(N_i)^a$ span $Z(\mathbf{n})$, we may now write the $Z(\mathbf{n})$-projection of \eqref{eq:div_J_0_raw} as:
\begin{align}
	& Z^{ab}\del_\mu j_b^{(0)\mu} = O(\varepsilon^2) \label{eq:div_J_0_Z} \\
	& \Rightarrow\quad Z^{ab}D n_b + n^a D_\mu\ell^\mu = O(\varepsilon^2).
\end{align}
Now consider the $Z^\bot(\mathbf{n})$-projection of \eqref{eq:div_J_0_raw}. It describes the vanishing of a first-order geometric scalar with a $(\dim\calG-\rank\calG)$-component charge index in $Z^\bot(\mathbf{n})$. Now, the only independent quantity of this type which can be constructed from $(h_{\mu\nu},\ell^\mu,s,n^a)$ is $Z^\bot_{ab}Dn^b$. Thus, unless the corrections $A_\mu^{(1)a}$ and $j^{(1)\mu}_a$ are such that eq. \eqref{eq:div_J_0_raw} becomes degenerate, it necessarily implies:
\begin{align}
  Z^\bot_{ab}Dn^b = O(\varepsilon^2). \label{eq:D_n_bot}
\end{align}
Conversely, suppose that non-degenerate first-order hydrodynamic equations exist (whether or not they are fully captured by the horizon equations \eqref{eq:GC}, \eqref{eq:YM_horizon}). The preceding discussion shows that these equations must contain eq. \eqref{eq:D_n_bot}. But then, that relation itself implies that the $f_{abc}$-terms in \eqref{eq:div_J_0_raw} must vanish. Indeed, $f_{abc}\mu^b\ell_\mu j^{(1)c\mu}$ and $f_{abc}n^b\ell^\mu A^{(1)c}_\mu$ are first-order scalars in $Z^\bot(\mathbf{n})$; if eq. \eqref{eq:D_n_bot} holds, then these quantities vanish at first order.

We conclude that the projected Yang-Mills equation \eqref{eq:div_J_0_raw} takes the form of the conservation law \eqref{eq:div_J_0}-\eqref{eq:D_q_0} in the non-abelian case as well.

\subsection{Gauss-Codazzi equations and energy-momentum conservation} \label{sec:ideal:GC}

The Gauss-Codazzi equation \eqref{eq:GC} reads at first order:
\begin{align}
	c^{(0)}_\mu \del_\nu S^{(0)\nu} + 2S^{(0)\nu}\del_{[\nu}c^{(0)}_{\mu]} = 16\pi j_a^{(0)\nu}\del_{[\mu}A^{(0)a}_{\nu]} + O(\varepsilon^2). \label{eq:GC_1}
\end{align}
Plugging in the zeroth-order values from section \ref{sec:homogeneous:thermo} and substituting $\del_\mu$ with $D_\mu$, the equation becomes:
\begin{align}
  \begin{split}
    & {-}8\pi\sqrt{-h}\left(D_\nu(Ts\ell_\mu\ell^\nu) + sD_\mu T\right) \\
    ={}& 8\pi\sqrt{-h}\left(D_\nu(\mu_a n^a\ell_\mu\ell^\nu) + n^a D_\mu \mu_a - \mu_a\ell_\mu D_\nu(n^a\ell^\mu)\right) + O(\varepsilon^2).
  \end{split}
\end{align}
In the last term on the RHS, we recognize the current divergence $\del_\nu j^{(0)\nu}_a$. Using the thermodynamic identities $\epsilon + p = Ts + \mu_a n^a$ and $dp = sdT + n^a d\mu_a$, we rearrange the equation to get:
\begin{align}
  -8\pi D_\nu T_\mu^{(0)\nu} = -8\pi\mu^a\ell_\mu D_\nu j^{(0)\nu}_a + O(\varepsilon^2), \label{eq:div_T_0_J}
\end{align}
where $T_\mu^{(0)\nu}$ is the stress-energy density defined in \eqref{eq:T_0}. Using eq. \eqref{eq:div_J_0} and dividing by $-8\pi$, we get an equation of hydrodynamic stress-energy conservation:
\begin{align}
  D_\nu T_\mu^{(0)\nu} = O(\varepsilon^2). \label{eq:div_T_0}
\end{align}
Together with the current conservation law \eqref{eq:div_J_0}, we now have a complete set of equations for ideal charged hydrodynamics. Note that these equations are independent of the structure constants $f_{abc}$ and the Chern-Simons coefficients $\beta_{abc}$; these represent interactions between adjacent regions of the brane/fluid, which will come into play only at the next order.

In components, eq. \eqref{eq:div_T_0} reads:
\begin{align}
  D\epsilon + (\epsilon + p)D_\mu\ell^\mu &= O(\varepsilon^2), \label{eq:D_epsilon_0} \\
  (\epsilon + p)D\ell^\mu + P^{\mu\nu}D_\nu p &= O(\varepsilon^2). \label{eq:a}
\end{align}
A convenient rewriting of eqs. \eqref{eq:D_q_0} and \eqref{eq:D_epsilon_0} is:
\begin{align}
  Ds &= -sD_\mu\ell^\mu + O(\varepsilon^2), \label{eq:D_s_0} \\
  D\frac{n_a}{s} &= O(\varepsilon^2). \label{eq:D_ns}
\end{align}
Eq. \eqref{eq:D_s_0} expresses the conservation of entropy in the evolution of an ideal fluid:
\begin{align}
  \del_\mu s^{(0)\mu} = O(\varepsilon^2).
\end{align}	
It can be derived more directly from the focusing equation \eqref{eq:focusing}, which reads at first order:
\begin{align}
	\kappa\theta = O(\varepsilon^2)\quad \Rightarrow\quad \frac{1}{v}\del_\mu S^\mu = \theta = O(\varepsilon^2).
\end{align}
This tells us that in the ideal approximation, the brane evolves in an area-preserving manner.

\section{First-order corrections and viscous hydrodynamics} \label{sec:viscous}

\subsection{Classification of first-order quantities} \label{sec:viscous:class}

We pause to consider the possible first-order quantities which can be constructed out of $(h_{\mu\nu},\ell^\mu,s,n^a)$. Some quantities are first-order for generic configurations, but become restricted to $O(\varepsilon^2)$ by the ideal constraint equations; here we discuss quantities which remain first-order after the ideal equations are imposed.

\subsubsection{Scalars} \label{sec:viscous:class:scalars}

Eqs. \eqref{eq:D_s_0}-\eqref{eq:D_ns} imply that $D_\mu\ell^\mu$ is the only independent first-order geometric scalar. Other commonly encountered geometric scalars have the form $Df$, where $f$ is some function of state (which may carry unspecified charge indices). Using \eqref{eq:D_s_0}-\eqref{eq:D_ns}, such quantities can be related to $D_\mu\ell^\mu$ as:
\begin{align}
  Df = -s\dds{f} D_\mu\ell^\mu + O(\varepsilon^2). \label{eq:D_f}
\end{align}
In particular, all first-order scalars with a single charge index have the form $f^a D_\mu\ell^\mu$, where $f^a$ is a charge vector constructed from $(s,n^a)$. As a result, the charge index of such quantities necessarily lies in $Z(\mathbf{n})$.

\subsubsection{Vectors} \label{sec:viscous:class:vectors}

Let us list the possible first-order vectors transverse to $\ell^\mu$. First, we have the acceleration $D\ell^\mu$ and the transverse gradients $P^{\mu\nu}\del_\nu f$ of thermodynamic functions $f$ (with possible charge indices). This list is redundant, since eq. \eqref{eq:a} relates $D\ell^\mu$ to $P^{\mu\nu}\del_\nu p$. Also, eq. \eqref{eq:D_ns} implies that $\del_\mu(n^a/s)$ is already transverse, with no need for a projector. Thus, a general transverse gradient can be written as:
\begin{align}
  P_\mu^\nu\del_\nu f = \dds{f} P_\mu^\nu\del_\nu s + s\ddn{f}{a} \del_\mu\frac{n^a}{s}\ . \label{eq:del_f}
\end{align}
Once axial quantities are allowed, we also have the vorticity vector density:
\begin{align}
  \omega^\mu \equiv \frac{1}{2}\epsilon^{\mu\nu\rho\sigma}\ell_\nu\del_\rho\ell_\sigma = \frac{1}{2}\epsilon^{\mu\nu\rho\sigma}\ell_\nu\omega_{\rho\sigma}\ , \label{eq:omega_vector}
\end{align}
where $\omega_{\mu\nu}$ is the vorticity tensor defined in \eqref{eq:omega}. Finally, we can get a first-order vector along $\ell^\mu$ by multiplying it with a first-order scalar.

We conclude that the most general first-order vector has the form:
\begin{align}
  V^{(1)\mu} = \tilde V\ell^\mu D_\nu\ell^\nu + V D^\mu s + V_a D^\mu\frac{n^a}{s} + \frac{V'}{\sqrt{-h}}\omega^\mu\ , \label{eq:V_1}
\end{align}
where the coefficients $\tilde V$, $V$, $V_a$ and $V'$ are $\calG$-covariant functions of $(s,n^a)$. To get a charged first-order vector, one should add charge indices to these coefficients in the obvious manner.

\subsubsection{Rank-2 tensors} \label{sec:viscous:class:tensors}

The only first-order traceless rank-2 tensors transverse to $\ell^\mu$ are the shear tensor $\pi_{\mu\nu}$ and the vorticity tensor $\omega_{\mu\nu}$:
\begin{align}
	\pi_{\mu\nu} &\equiv P_\mu^\rho P_\nu^\sigma D_{(\rho}\ell_{\sigma)} - \frac{1}{3}P_{\mu\nu}D_\rho\ell^\rho\ , \label{eq:pi} \\
	\omega_{\mu\nu} &\equiv P_\mu^\rho P_\nu^\sigma D_{[\rho}\ell_{\sigma]} = -3\ell^\rho\ell_{[\rho}\del_\mu\ell_{\nu]}\ . \label{eq:omega}
\end{align}
Other first-order rank-2 tensors can be constructed from the tensor product of $\ell^\mu$ with first-order vectors, or of $P_{\mu\nu}$ with first-order scalars.

\subsection{Corrections to the horizon ansatz} \label{sec:viscous:corrections}

Before we write down the second-order constraint equations, we must consider the possible first-order corrections to the horizon ansatz \eqref{eq:grav_0}-\eqref{eq:charge_0}. These corrections can be restricted by a first-order fixing of variables. We consider $(\ell^\mu,s,n^a)$ as the basic variables in \eqref{eq:grav_0}-\eqref{eq:charge_0}, with $T$ and $\mu_a$ derived from $(s,n^a)$ through the equation of state. We fix the direction of the entropy velocity $\ell^\mu$ and the magnitude of the entropy density $s$ by requiring that eq. \eqref{eq:S_0} remains without corrections, i.e.:
\begin{align}
  S^{(1)\mu} = 0; \quad v^{(1)} = 0\ .
\end{align}
Due to this condition, the correction to $\gamma_{\mu\nu}$ must be transverse to $\ell^\mu$ and traceless with respect to $h_{\mu\nu}$. We will find that the precise form of $\gamma^{(1)}_{\mu\nu}$ is otherwise irrelevant to the constraint equations. For completeness, we note that this correction must take the form:
\begin{align}
  \gamma^{(1)}_{\mu\nu} = \alpha\pi_{\mu\nu}\ ,
\end{align}
with $\alpha$ some function of $(s,n^a)$.

Let us now turn to the $\gamma_{\mu\nu}$-independent components $c_\mu$ of the extrinsic curvature. For the definition of $c_\mu$, we still use the auxiliary matrix \eqref{eq:G_specific}. The most general correction to $c_\mu$ reads:
\begin{align}
  c_\mu^{(1)} = \tilde c\ell_\mu D_\nu\ell^\nu + c D_\mu s + c_a D_\mu\frac{n^a}{s} + \frac{c'}{\sqrt{-h}}\omega_\mu\ , \label{eq:c_1}
\end{align}
where $\tilde c$, $c$, $c_a$ and $c'$ are some functions of $(s,n^a)$. The requirement for the constraint equations to take the form of conservation laws will place restrictions on these functions. Otherwise, they will have no impact on the hydrodynamic transport coefficients.

We fix the meaning of the charge density $n^a$ by requiring the relation $\ell_\mu j_a^\mu = -\sqrt{-h}n_a$ to hold without corrections:
\begin{align}
  \ell_\mu j_a^{(1)\mu} = 0\ . \label{eq:n_condition}
\end{align}
As we will see, this condition fixes $n^a$ as the longitudinal component $-\ell_\mu J_a^\mu/\sqrt{-h}$ of the conserved hydrodynamic charge current. As implied by \eqref{eq:j_F}, all the remaining first-order corrections to $j^\mu_a$ are determined by the horizon gauge field $A_\mu^a$. The general correction to $A_\mu^a$ takes the form:
\begin{align}
  A_{a\mu}^{(1)} = \tilde A_a\ell_\mu D_\nu\ell^\nu + A_a D_\mu s + A_{ab} D_\mu\frac{n^b}{s} + \frac{A'_a}{\sqrt{-h}}\omega_\mu\ , \label{eq:A_1}
\end{align}
where $\tilde A_a$, $A_a$, $A_{ab}$ and $A'_a$ are some functions of $(s,n^a)$. Again, the requirement for a conservation-law form of the equations places restrictions on the form of these functions. In the abelian case, they otherwise have no impact on the resulting transport coefficients. In the non-abelian case, however, the $Z^\bot(\mathbf{n})$-components of $A_{ab}$ play a crucial role in determining the $Z^\bot(\mathbf{n})$-piece of the conductivity matrix.

\subsection{Current conservation} \label{sec:viscous:current}

In this subsection, we analyze the second-order terms in the projected Yang-Mills equation \eqref{eq:YM_horizon}. The abelian case is simpler, and we will consider it first.

\subsubsection{Abelian case}

When $f_{abc} = 0$, the horizon's intrinsic field strength $F^a_{\mu\nu}$ can be written as:
\begin{align}
  F^a_{\mu\nu} = 2\del_{[\mu}A^{(0)a}_{\nu]} + O(\varepsilon^2) = 2D_{[\nu}(\mu^a\ell_{\mu]}) + O(\varepsilon^2). \label{eq:F_1_abel}
\end{align}
To find $j^{(1)\mu}_a$, we write down the first-order terms of eq. \eqref{eq:j_F}:
\begin{align}
  \gamma^{(1)}_{\mu\nu}j^{(0)a\nu} + \gamma^{(0)}_{\mu\nu}j^{(1)a\nu} = \frac{1}{4\pi}S^{(0)\nu} F^{(1)a}_{\nu\mu}\ .
\end{align}
The first term vanishes due to $\gamma^{(1)}_{\mu\nu}\ell^\nu = 0$. The other terms evaluate as:
\begin{align}
  (4s)^{2/3}P_{\mu\nu}j_a^{(1)\nu} = -\frac{\sqrt{-h}s}{\pi}(P^\nu_\mu D_\nu\mu_a + \mu_a D\ell_\mu).
\end{align}
Together with eq. \eqref{eq:n_condition}, this gives:
\begin{align}
  j_a^{(1)\mu} = -\frac{\sqrt{-h}s^{1/3}}{2^{4/3}\pi}(P^\nu_\mu D_\nu\mu_a + \mu_a D\ell_\mu). \label{eq:j_1_abelian}
\end{align}
From equation \eqref{eq:a} and the identity $dp = sdT + n^a d\mu_a$, we derive an identity of ideal hydrodynamics:
\begin{align}
  P_\mu^\nu D_\nu\mu_a + \mu_a D\ell_\mu = T\left(\delta_a^b - \frac{\mu_a n^b}{\epsilon + p}\right) P_\mu^\nu D_\nu\frac{\mu_b}{T} + O(\varepsilon^2), \label{eq:mu_muT}
\end{align}
which enables us to rewrite \eqref{eq:j_1_abelian} as:
\begin{align}
  j_a^{(1)\mu} &= -\sqrt{-h}T\bar\sigma_{ab} P_\mu^\nu D_\nu\frac{\mu^b}{T}\ , \label{eq:j_1_abelian2} \\
  \bar\sigma_{ab} &\equiv \frac{s^{1/3}}{2^{4/3}\pi}\left(\delta_{ab} - \frac{\mu_a n_b}{\epsilon + p}\right). \label{eq:sigma_abel}
\end{align}
For the Chern-Simons current \eqref{eq:j_CS}, we have:
\begin{align}
  \Jcs^{(1)a\mu} = -\frac{\beta_{abc}}{\pi}\epsilon^{\mu\nu\rho\sigma} A^{(0)b}_\nu \del_\rho A^{(0)c}_\sigma
    = -\frac{2}{\pi}\beta_{abc}\mu^b\mu^c\omega^\mu\ . \label{eq:Jcs_1}
\end{align}

To second order accuracy, the horizon Yang-Mills equation \eqref{eq:YM_horizon} now reads:
\begin{align}
  \del_\mu J_a^\mu = O(\varepsilon^3),
\end{align}
where the conserved current is given by:
\begin{align}
  J^{a\mu} &= j^{(0)a\mu} + j^{(1)a\mu} + \Jcs^{(1)a\mu} + O(\varepsilon^2) \notag\\
    &= \sqrt{-h}\left(n^a\ell^\mu - T\bar\sigma^{ab} P^{\mu\nu} D_\nu\frac{\mu_b}{T}\right) - \frac{2}{\pi}\beta^{abc}\mu_b\mu_c\omega^\mu + O(\varepsilon^2). \label{eq:div_J_1_abel}
\end{align}

\subsubsection{Non-abelian case}

The non-abelian case requires a more elaborate argument. The first-order field strength $F^{(1)a}_{\mu\nu}$ now receives an additional contribution:
\begin{align}
  2f^a{}_{bc}A^{(0)b}_{[\mu}A^{(1)c}_{\nu]} = -2f^a{}_{bc}\mu^b\ell_{[\mu}A^{(1)c}_{\nu]}\ .
\end{align}
The factor of $f^a{}_{bc}\mu^b$ singles out the $Z^\bot(\mathbf{n})$-components of $A^{(1)c}_\nu$. Of all the terms in \eqref{eq:A_1}, this leaves only the $A_{ab}$-term; the other coefficients are charge vectors built out of $(s,n^a)$, and therefore lie in $Z(\mathbf{n})$. Furthermore, we may rewrite the $A_{ab}$-term as:
\begin{align}
  A_{ab} D_\mu\frac{n^b}{s} &= \bar A_{ab} D_\mu\mu^b + \dots \\
  \bar A_{ab} &\equiv \frac{A_{ac}}{s} \dd{n^c}{\mu^b}{T},
\end{align}
where the ellipses indicate terms consisting of (zeroth-order) charge vectors in $Z(\mathbf{n})$ multiplied by (first-order) charge scalars. These terms will again be annihilated by $f^a{}_{bc}\mu^b$. In summary, we may write:
\begin{align}
  F_{a\mu\nu} = 2\left(Z_{ab}D_{[\nu}(\mu^b\ell_{\mu]}) + j^\bot_{ab}\ell_{[\mu}D_{\nu]}\mu^b\right) + O(\varepsilon^2), \label{eq:F_1}
\end{align}
where $j^\bot_{ab}$ is the following matrix, whose indices lie in $Z^\bot(\mathbf{n})$:
\begin{align}
  j^\bot_{ab} = Z^\bot_{ab} - f_{acd}\mu^c\bar A^d{}_b\ . \label{eq:j_bot}
\end{align}
Analogously to \eqref{eq:j_1_abelian2} in the abelian case, we can now find the first-order correction to $j^\mu_a$ as:
\begin{align}
  j_a^{(1)\mu} = -\sqrt{-h}T\left(\bar\sigma^{||}_{ab} + \frac{s^{1/3}}{2^{4/3}\pi}j^\bot_{ab}\right)P^{\mu\nu} D_\nu\frac{\mu^b}{T}\ , \label{eq:j_1}
\end{align}
where we used $j^\bot_{ab}\mu^b = 0$ to manipulate the $j^\bot_{ab}$-term, and the matrix $\bar\sigma^{||}_{ab}$ with indices in $Z(\mathbf{n})$ is defined similarly to \eqref{eq:sigma_abel}:
\begin{align}
  \bar\sigma^{||}_{ab} = \frac{s^{1/3}}{2^{4/3}\pi}\left(Z_{ab} - \frac{\mu_a n_b}{\epsilon + p}\right). \label{eq:sigma_Z_bar}
\end{align}

Turning to the Chern-Simons current \eqref{eq:j_CS}, it's easy to see that the first-order expression \eqref{eq:Jcs_1} is valid in the non-abelian case as well: there is no contribution from $f_{abc}$-terms, because $A^{(0)a}_\mu \sim \ell_\mu$. Examining eq. \eqref{eq:Jcs_1}, we see that $\beta_{abc}\mu^b\mu^c$ is a gauge vector covariantly constructed out of $(n^a,s)$. It follows that $\beta_{abc}\mu^b\mu^c$, and therefore $\Jcs^{(1)a\mu}$ itself, lie in $Z(\mathbf{n})$.

We are now ready to write down the second-order terms in the LHS of the restricted Yang-Mills equation \eqref{eq:YM_horizon}:
\begin{align}
  \begin{split}
	& \del_\mu(j^{(1)a\mu} + \Jcs^{(1)a\mu})
			+ f^a{}_{bc}\left(A^{(1)b}_\mu(j^{(1)c\mu} + \Jcs^{(1)c\mu}) + A^{(0)b}_\mu(j^{(2)c\mu} + \Jcs^{(2)c\mu}) + A^{(2)b}_\mu j^{(0)c\mu}\right) \\
	& 	+ \frac{\beta^a{}_{bc}}{\pi}\epsilon^{\mu\nu\rho\sigma}f^b{}_{de}A^{(0)d}_\mu A^{(1)e}_\nu F^{(1)c}_{\rho\sigma} \\
	={}& \del_\mu(j^{(1)a\mu} + \Jcs^{(1)a\mu}) + f^a{}_{bc} A^{(1)b}_\mu j^{(1)c\mu} \\
	&	+ f^a{}_{bc} A^{(1)b}_\mu\Jcs^{(1)c\mu} - f^a{}_{bc}\mu^b\ell_\mu(j^{(2)c\mu} + \Jcs^{(2)c\mu})
			+ \sqrt{-h}f^a{}_{bc}\ell^\mu A^{(2)b}_\mu n^c \\
	&	+ \frac{2}{\pi}\beta^a{}_{bc}\mu^c f^{bde}\mu_d \epsilon^{\mu\nu\rho\sigma}\ell_\mu A^{(1)e}_\nu \omega_{\rho\sigma}\ . \label{eq:YM_horizon_2}
  \end{split}
\end{align}
Let us contract \eqref{eq:YM_horizon_2} with a charge vector $(N_i)_a$ covariantly constructed from $n^a$ (recall that such vectors span $Z(\mathbf{n})$). We find that all the terms in the last two lines vanish:
\begin{itemize}
	\item In the $\Jcs^{(1)c\mu}$-term, we have $f^a{}_{bc}(N_i)_a\Jcs^{(1)c\mu} = 0$, because $(N_i)_a,\Jcs^{(1)c\mu} \in Z(\mathbf{n})$.
	\item In the $(j^{(2)c\mu} + \Jcs^{(2)c\mu})$-term, we have $f^a{}_{bc}(N_i)_a \mu^b = 0$, because $(N_i)_a,\mu^b \in Z(\mathbf{n})$.
	\item In the $A^{(2)b}_\mu$-term, we have $f^a{}_{bc}(N_i)^a n^c = 0$, because $(N_i)_a \in Z(\mathbf{n})$.
	\item In the $\beta^a{}_{bc}$-term, we have $f^{bde}\mu_d\,\beta^a{}_{bc}(N_i)_a\mu^c = 0$, because $\beta^a{}_{bc}(N_i)_a\mu^c \in Z(\mathbf{n})$, as a charge vector covariantly constructed from $(n^a,s)$.
\end{itemize}
In the contraction of the $f^a{}_{bc} A^{(1)b}_\mu j^{(1)c\mu}$-term with $(N_i)_a$, only the $Z^\bot(\mathbf{n})$-components of $A^{(1)b}_\mu$ and $j^{(1)c\mu}$ survive. We get:
\begin{align}
	(N_i)_a f^a{}_{bc} A^{(1)b}_\mu j^{(1)c\mu}
		= -\frac{\sqrt{-h}s^{1/3}T}{2^{4/3}\pi}f^{abc}(N_i)_a \bar A_{bd} j^\bot_{ce} P^{\mu\nu}D_\nu\mu^d D_\mu\frac{\mu^e}{T}\ . \label{eq:Q_f_A_j}
\end{align}
Generically, $f_{abc}(N_i)^c$ is an invertible matrix over $Z^\bot(\mathbf{n})$. We may therefore denote:
\begin{align}
	j^\bot_{ab} = k_a{}^c f_{cbd}(N_i)^d = f_{acd}(N_i)^d k^c{}_b\ ,
\end{align}
where we used the matrix commutation of $k_{ab}$ and $f_{abc}(N_i)^c$ (Proposition \ref{prop:commutation}). Expression \eqref{eq:Q_f_A_j} now becomes:
\begin{align}
  \begin{split}
	& {-}\frac{\sqrt{-h}s^{1/3}T}{2^{4/3}\pi}f^{abc}(N_i)_a \bar A_{bd} k_c{}^f f_{feg}(N_i)^g P^{\mu\nu}D_\nu\mu^d D_\mu\frac{\mu^e}{T} \\
	={}& {-}(N_i)^g D_\mu\left(\frac{\sqrt{-h}s^{1/3}}{2^{4/3}\pi}f^{abc}(N_i)_a \bar A_{bd} k_c{}^f f_{feg}\mu^e P^{\mu\nu}D_\nu\mu^d\right) \\
	={}& {-}(N_i)^g D_\mu\left(\frac{\sqrt{-h}s^{1/3}}{2^{4/3}\pi}\bar A_{bd} j^\bot_{bf} f_{feg}\mu^e P^{\mu\nu}D_\nu\mu^d\right) \\
	={}& {-}(N_i)^g D_\mu\left(\frac{\sqrt{-h}s^{1/3}}{2^{4/3}\pi}\bar A_{bd} f_{bef}\mu^e j^\bot_{fg} P^{\mu\nu}D_\nu\mu^d\right) \\
	={}& (N_i)^g D_\mu\left(\frac{\sqrt{-h}s^{1/3}}{2^{4/3}\pi}(Z^\bot_{fd} - j^\bot_{fd}) j^\bot_{fg} P^{\mu\nu}D_\nu\mu^d\right) \\
	={}& (N_i)^a D_\mu\left(\frac{\sqrt{-h}s^{1/3}T}{2^{4/3}\pi}(j^\bot_{ba} - j^\bot_{ca} j^\bot_{cb}) P^{\mu\nu}D_\nu\frac{\mu^b}{T}\right).
  \end{split}
\end{align}
Thus, the full contraction of \eqref{eq:YM_horizon_2} with $(N_i)^a$ reads:
\begin{align}
	(N_i)^a D_\mu\left(-\sqrt{-h}T\left(\bar\sigma^{||}_{ab} + \frac{s^{1/3}}{2^{4/3}\pi}(2j^\bot_{[ab]} + j^\bot_{ca}j^\bot_{cb})\right)
	  P^{\mu\nu}D_\nu\frac{\mu^b}{T} + \Jcs^{(1)a\mu}\right). \label{eq:Q_YM_2}
\end{align}
The $j^\bot_{[ab]}$-term is actually redundant. Indeed, let $X_{ab} = X_{[ab]}$ be an antisymmetric matrix with indices in $Z^\bot(\mathbf{n})$. As our spanning set $(N_i)_a$ of $Z(\mathbf{n})$, we may choose the derivatives $\del N_i/\del \mu^a$ of a set of scalar functions of $\mu^a$. Then $(N_i)_{ab} \equiv \del (N_i)_a/\del \mu^b$ is symmetric\footnote{Since the $(N_i)_a$ span $Z(\mathbf{n})$, this symmetry property follows for \emph{every} charge vector constructed from $\mu^a$ (or $n^a$).}. Furthermore, according to Proposition \ref{prop:commutation}, $X_{ab}$ and $(N_i)_{ab}$ commute as matrices, so their matrix product is antisymmetric. We then find:
\begin{align}
  \begin{split}
	(N_i)^a D_\mu\left(X_{ab} P^{\mu\nu} D_\nu\frac{\mu^b}{T}\right) &= (N_i)^a D_\mu\left(\frac{1}{T}X_{ab}P^{\mu\nu} D_\nu\mu^b\right)
		= -\frac{1}{T}X_{ab} P^{\mu\nu} D_\mu(N_i)^a D_\nu\mu^b \\
		&= -\frac{1}{T}X_{ab}(N_i)^a_c P^{\mu\nu} D_\mu\mu^c D_\nu\mu^b = 0\ ,
  \end{split}
\end{align}
where we used the relations $X_{ab}\mu^b = X_{ab}(N_i)^b = 0$ and the antisymmetry of $X_{ab}(N_i)^a_c$. We conclude that the $j^\bot_{[ab]}$-term in \eqref{eq:Q_YM_2} may be replaced with any other antisymmetric matrix over $Z^\bot(\mathbf{n})$. We can then write \eqref{eq:Q_YM_2} as:
\begin{align}
	(N_i)^a D_\mu\left(-\sqrt{-h}T(\bar\sigma^{||}_{ab} + \sigma^\bot_{ab}) P^{\mu\nu}D_\nu\frac{\mu^b}{T} + \Jcs^{(1)a\mu}\right), \label{eq:Q_YM_2_general}
\end{align}
where $\sigma^\bot_{ab}$ is a function of $(s,n^a)$ with indices in $Z^\bot(\mathbf{q})$ whose symmetric piece is given by
\begin{align}
	\sigma^\bot_{(ab)} = \frac{s^{1/3}}{2^{4/3}\pi}j^\bot_{ca}j^\bot_{cb}\ , \label{eq:sigma_bot_symm}
\end{align}
and the antisymmetric piece $\sigma^\bot_{[ab]}$ is otherwise arbitrary. Eq. \eqref{eq:sigma_bot_symm} implies that $\sigma^\bot_{(ab)}$ is positive semi-definite. As we will see, this property is related to the second law of thermodynamics.

Since the $(N_i)^a$ span $Z(\mathbf{n})$, we conclude from \eqref{eq:Q_YM_2_general} that the $Z(\mathbf{n})$-projection of the second-order terms \eqref{eq:YM_horizon_2} is the $Z(\mathbf{n})$-projection of a divergence. Combining this with the results of the previous order, we get:
\begin{align}
	Z^{ab}\del_\mu J_b^\mu = Z^{ab}\del_\mu(J_b^{(0)\mu} + J_b^{(1)\mu}) = O(\varepsilon^3), \label{eq:div_J_1_Z}
\end{align}
where $J_a^{(1)\mu}$ is given by
\begin{align}
  J_a^{(1)\mu} &= - T\bar\sigma_{ab}P^{\mu\nu}D_\nu\frac{\mu^b}{T} - \frac{2}{\pi}\beta_{abc}\mu^b\mu^c\omega^\mu + O(\varepsilon^2) \label{eq:J_1}
\end{align}
with $\bar\sigma_{ab} \equiv \bar\sigma^{||}_{ab} + \sigma^\bot_{ab}$.

Let us now return to the full second-order expression \eqref{eq:YM_horizon_2}. We see that it depends on second-order corrections to the horizon ansatz. Specifically, this dependence is contained in the terms $f^a{}_{bc}(A^{(2)b}_\mu j^{(0)c\mu} + A^{(0)b}_\mu j^{(2)c\mu})$. At first sight, we may be concerned that this quantity is sensitive to the second-order fixing of variables. Reassuringly, this isn't so. Indeed, consider a second-order redefinition of the hydrodynamic variables $(h_{\mu\nu},\ell^\mu,s,n^a)$. A redefinition of variables means that the value of geometric and gauge quantities such as $f^a{}_{bc}A^b_\mu j^{c\mu}$ must remain unchanged, while their functional dependence on $(h_{\mu\nu},\ell^\mu,s,n^a)$ may be altered. In particular, the contribution to $f^a{}_{bc}(A^{(2)b}_\mu j^{(0)c\mu} + A^{(0)b}_\mu j^{(2)c\mu})$ due to the redefinition must equal minus the change in the value of $f^a{}_{bc}A^{(0)b}_\mu j^{(0)c\mu} = \sqrt{-h}f^a{}_{bc}\mu^b n^c$. But $f^a{}_{bc}\mu^b n^c$ always vanishes regardless of the variable-fixing, since $\mu^a$ is always a covariant function of $(s,n^a)$. We conclude that expression \eqref{eq:YM_horizon_2} is insensitive to the second-order fixing of variables.

Now, since the matrices $f^a{}_{bc} n^c$ and $f^a{}_{bc}\mu^b$ are invertible over $Z^\bot(\mathbf{n})$, we see that appropriate choices of $A^{(2)a}_\mu$ and $j_a^{(2)\mu}$ can make the $Z^\bot(\mathbf{n})$-projection of \eqref{eq:YM_horizon_2} equal \emph{anything}. In particular, we can make it equal $Z^\bot_{ab}\del_\mu J^{(1)b\mu}$, with any choice of $\sigma^\bot_{[ab]}$ in \eqref{eq:J_1}. Then eq. \eqref{eq:div_J_1_Z} is upgraded into a full conservation law:
\begin{align}
	\del_\mu J_a^\mu = O(\varepsilon^3). \label{eq:div_J_1}
\end{align}
Thus, for an arbitrary first-order correction $A^{(1)a}_\mu$ to the gauge potential, there exist appropriate second-order corrections $(A_\mu^{(2)a}, j^{(2)\mu}_a)$ for which the horizon Yang-Mills equation takes the form of a conservation law. The first-order correction determines the symmetric part $\sigma^\bot_{(ab)}$ of the $Z^\bot(\mathbf{n})$-piece of the conductivity matrix, while the second-order corrections determine its antisymmetric part $\sigma^\bot_{[ab]}$.

One of the allowed sets of first-order and second-order corrections corresponds to the non-abelian AdS black brane with vanishing gauge fields at $r \rightarrow \infty$. Unfortunately, we cannot find the relevant corrections on the horizon without solving the radial equations. Thus, we do not obtain the specific form of $\sigma^\bot_{ab}$. However, our analysis provides two non-trivial relations between the gauge fields on the horizon and on the AdS boundary:
\begin{itemize}
	\item $\sigma^\bot_{(ab)}$ appears in the conserved current \eqref{eq:J_1}, which should correspond to the boundary quantity \eqref{eq:J_boundary}. On the other hand, $\sigma^\bot_{(ab)}$ is defined in \eqref{eq:sigma_bot_symm} in terms of $j^\bot_{ab}$, which is derived from the gauge potential $A^{(1)a}_\mu$ \emph{on the horizon}. This provides a first-order relation between horizon and boundary fields. In the simplest non-abelian case $\calG = SU(2)$, we can check this relation against the results of the bulk calculation in \cite{Torabian:2009qk}.
In the notation of \cite{Torabian:2009qk}, the first-order relation we find reads:
\begin{align}
	\lim_{r\rightarrow\infty}r^2 g^{(1)} = \frac{\sqrt{3}}{4R}\left.\left(\frac{1}{q^2} + 2f^{(1)} + q^2\left((f^{(1)})^2 + q^2(g^{(1)})^2\right)\right)\right|_{\calH}\ , \label{eq:SU2_relation}
\end{align}
where $f^{(1)}$ and $g^{(1)}$ are the coefficients of the two non-abelian terms in $A^a_\mu$, and the RHS is evaluated at the horizon.
 This is a scalar equation, because for an $SU(2)$ group, $\sigma^\bot_{(ab)}$ has a single component $\sim(q^2\delta_{ab} - q_a q_b)$. Following the numerical prescription of \cite{Torabian:2009qk} for solving the radial differential equations, we have calculated $f^{(1)}(r)$ and $g^{(1)}(r)$ for several values of $q^2/R^6$ in the range $0.01-1.7$, and found that \eqref{eq:SU2_relation} holds within the numerical accuracy of $\sim 1\%$. We view this as evidence for the correct identification of the conserved current \eqref{eq:J_1} with the AdS/CFT boundary current \eqref{eq:J_boundary}.
	\item The condition that the $Z^\bot(\mathbf{q})$-projection of \eqref{eq:YM_horizon_2} should equal $Z^\bot_{ab}\del_\mu J^{(1)b\mu}$ provides another relation, at second order, between horizon fields ($A^{(1)a}_\mu$, $A^{(2)a}_\mu$) and boundary fields ($J_a^{(1)\mu}$, with its $\sigma^\bot_{(ab)}$ and $\sigma^\bot_{[ab]}$ terms).
\end{itemize}

\subsection{Gauss-Codazzi equations and energy-momentum conservation} \label{sec:viscous:GC}

In this subsection, we analyze the second-order Gauss-Codazzi equation. The non-abelian case doesn't lead to significant complications, so we will handle it straight away.

The horizon's first-order shear/expansion tensor can be derived directly from the zeroth-order metric via \eqref{eq:theta_def}:
\begin{align}
  \begin{split}
	\theta_{\mu\nu} &= \frac{1}{2}\Liebold{\ell}\gamma^{(0)}_{\mu\nu} + O(\varepsilon^2)
		= \frac{1}{2}\ell^\rho D_\rho\gamma^{(0)}_{\mu\nu} + \gamma^{(0)}_{\rho(\mu} D_{\nu)}\ell^\rho + O(\varepsilon^2) \\
		&= (4s)^{2/3}\left(\pi_{\mu\nu} + \frac{1}{3s}(Ds + sD_\rho\ell^\rho)P_{\mu\nu}\right) + O(\varepsilon^2)
		= (4s)^{2/3}\pi_{\mu\nu} + O(\varepsilon^2). \label{eq:theta_1}
  \end{split}
\end{align}
In the second equality, we wrote the Lie derivative in terms of the connection $D_\mu$. In the last, we used the ideal equation \eqref{eq:D_s_0}. Decomposing the result with \eqref{eq:theta_decompose}-\eqref{eq:theta_trace}, we get:
\begin{align}
	\theta^{(1)} = 0; \quad \sigma^{(H)(1)}_{\mu\nu} = (4s)^{2/3}\pi_{\mu\nu}\ . \label{eq:theta_sigma_1}
\end{align}
The result for $\theta^{(1)}$ was already known from the ideal equations.

Let us now evaluate the second-order terms in the Gauss-Codazzi equation \eqref{eq:GC}. For the LHS of \eqref{eq:GC} we get, using our choice $\lambda = 1/(4s)^{2/3}$:
\begin{align}
	 D^{(G)}_\nu\left(\frac{v^{(0)}}{(4s)^{2/3}}\theta^{(1)}_{\mu\rho}(G^{-1})^{\rho\nu}\right) + 2S^{(0)\nu}\del_{[\nu}c^{(1)}_{\mu]}\ ,
\end{align}
where we can use the zeroth-order expression \eqref{eq:G_0} for $G_{\mu\nu}$, substituting $(G^{-1})^{\rho\nu}$ with $h^{\rho\nu}$ and $D^{(G)}_\nu$ with $D_\nu$. This gives:
\begin{align}
	D_\nu\left(\frac{v^{(0)}}{(4s)^{2/3}}\theta^{(1)\nu}_\mu\right) + 2S^{(0)\nu}\del_{[\nu}c^{(1)}_{\mu]}
		= D_\nu\left(4\sqrt{-h}s\pi_\mu^\nu\right) + 8\sqrt{-h}s\ell^\nu\del_{[\nu}c^{(1)}_{\mu]}\ . \label{eq:GC_2_LHS}
\end{align}
We now turn to the RHS of \eqref{eq:GC}. At second order, it reads:
\begin{align}
	8\pi(F^{(2)a}_{\mu\nu}j_a^{(0)\nu} + F^{(1)a}_{\mu\nu}j_a^{(1)\nu}). \label{eq:GC_2_RHS}
\end{align}
The first term can be expanded as:
\begin{align}
  \begin{split}
	8\pi F^{(2)a}_{\mu\nu}j_a^{(0)\nu}
		&= 8\pi\sqrt{-h}n_a\ell^\nu\left(2\del_{[\mu}A^{(1)a}_{\nu]} + f^a{}_{bc}(2A^{(0)b}_{[\mu}A^{(2)c}_{\nu]} + A^{(1)}_{b\mu}A^{(1)}_{c\nu})\right) \\
		&= 16\pi\sqrt{-h}n_a\ell^\nu\del_{[\mu}A^{(1)a}_{\nu]} + O(\varepsilon^3), \label{eq:del_A_1}
  \end{split}
\end{align}
where we used the fact that both $A^{(0)a}_\mu$ and the first-order geometric scalar $\ell^\nu A^{(1)}_{c\nu}$ lie in $Z(\mathbf{n})$. The second term in \eqref{eq:GC_2_RHS} reads:
\begin{align}
  \begin{split}
   8\pi F^{(1)a}_{\mu\nu}j_a^{(1)\nu} ={}& {-}8\pi\sqrt{-h}\left(T\bar\sigma^{||}_{ab} P^{\nu\rho} D_\rho\frac{\mu^b}{T}\left(D_\nu(\mu^a\ell_\mu) - \mu^a D_\mu\ell_\nu\right)\right. \\
    & \left.\vphantom{D_\rho\frac{\mu^b}{T}} + \ell_\mu\sigma^\bot_{(ab)}P^{\nu\rho} D_\nu\mu^a D_\rho\mu^b\right), \label{eq:F_1_j_1}
  \end{split}
\end{align}
where we used eqs. \eqref{eq:F_1}, \eqref{eq:j_1}, \eqref{eq:omega} and \eqref{eq:sigma_bot_symm}, and the fact that $\sigma^\bot_{(ab)}\mu^b = 0$.

There is another second-order correction that must be taken into account. At first order, we brought the Einstein-Gauss-Codazzi equation to the form \eqref{eq:div_T_0_J}. The RHS of that equation vanished due to the first-order current conservation. At second order, this is no longer true. Instead, we have:
\begin{align}
  \begin{split}
	-8\pi\mu^a\ell_\mu D_\nu J_a^{(0)\nu} ={}& {-}8\pi\mu^a\ell_\mu D_\nu(J_a^\nu - J_a^{(1)\nu}) + O(\varepsilon^3) \\
	  ={}& {-}8\pi\mu^a\ell_\mu D_\nu J_a^\nu
	  - 8\pi\ell_\mu\left(\mu^a D_\nu\left(\sqrt{-h}T\bar\sigma^{||}_{ab} P^{\nu\rho}D_\rho\frac{\mu^b}{T} + \frac{2}{\pi}\beta_{abc}\mu^b\mu^c\omega^\nu\right)\right. \\
	  & - \left. \vphantom{\left(\frac{\mu^b}{T}\right)} \sqrt{-h}\sigma^\bot_{ab}P^{\nu\rho}D_\nu\mu^a D_\rho\mu^b\right) + O(\varepsilon^3), \label{eq:div_J_1_GC}
  \end{split}
\end{align}
where we used eq. \eqref{eq:J_1} and applied the Leibnitz rule on the $\sigma^\bot_{ab}$-term. We see that the $\sigma^\bot_{ab}$-terms in \eqref{eq:F_1_j_1} and \eqref{eq:div_J_1_GC} cancel each other. The $\bar\sigma^{||}_{ab}$-terms combine to give:
\begin{align}
  -8\pi D_\nu\left(\sqrt{-h}T\mu^a\bar\sigma^{||}_{ab}\ell_\mu P^{\nu\rho}D_\rho\frac{\mu^b}{T}\right) + 8\pi\sqrt{-h}T\mu^a\bar\sigma^{||}_{ab} D^\nu\frac{\mu^b}{T} D_\mu\ell_\nu\ . \label{eq:sigma_Z_terms}
\end{align}
We must now use the Leibnitz rule repeatedly to manipulate the second term of \eqref{eq:sigma_Z_terms}, keeping the products $T\mu^a\bar\sigma^{||}_{ab}$ and $\mu^b/T$ intact. The expression becomes:
\begin{align}
  \begin{split}
    & -16\pi D^\nu\left(\sqrt{-h}T\mu^a\bar\sigma^{||}_{ab}\ell_{(\mu} P_{\nu)}^\rho D_\rho\frac{\mu^b}{T}\right) \\
    & + 8\pi\sqrt{-h}\left(D_\nu\left(P_\mu^\nu T\mu^a\bar\sigma^{||}_{ab} D\frac{\mu^b}{T}\right) - D\frac{\mu^b}{T}D_\mu(T\mu^a\bar\sigma^{||}_{ab})
      + D_\nu(T\mu^a\bar\sigma^{||}_{ab}\ell^\nu)D_\mu\frac{\mu^b}{T}\right).
  \end{split}
\end{align}
The first line is the divergence of a symmetric tensor; it can be used as part of a stress-energy conservation law. The terms in the second line we intend to cancel with other contributions. For that purpose, we decompose the gradients in the last two terms into combinations of $\del_\mu s$ and $\del_\mu(n^a/s)$, and use the ideal equation \eqref{eq:D_f}. The second line then reads:
\begin{align}
  \begin{split}
    & 8\pi\sqrt{-h}\left(D_\nu\left(P_\mu^\nu T\mu^a\bar\sigma^{||}_{ab} D\frac{\mu^b}{T}\right) + T\mu^a\bar\sigma^{||}_{ab} D_\nu\ell^\nu \dds{(\mu^b/T)} D_\mu s\right. \\
    &\ + \left.s\left(D_\nu(T\mu^a\bar\sigma^{||}_{ab}\ell^\nu)\ddn{(\mu^b/T)}{c} - D\frac{\mu^b}{T}\ddn{(T\mu^a\bar\sigma^{||}_{ab})}{c}\right) D_\mu\frac{n^c}{s}\right). \label{eq:sigma_terms}
  \end{split}
\end{align}

We now turn to the Chern-Simons term in \eqref{eq:div_J_1_GC}. To approach it, we will need two identities concerning the vorticity:
\begin{align}
  D_\mu\omega^\mu &= 2\omega^\mu D\ell_\mu\ , \\
  \omega^\nu\omega_{\nu\mu} &= 0\ . \label{eq:omega_omega}
\end{align}
The first of these is easy to derive from the definition \eqref{eq:omega_vector} and the four-dimensionality of the horizon. To derive the second, note that $\omega_{\mu\nu}$ is effectively three-dimensional, as it is transverse to $\ell^\mu$. This implies that $\omega_{[\mu\nu}\omega_{\rho\sigma]} = 0$ identically. Eq. \eqref{eq:omega_omega} then follows from the definition \eqref{eq:omega_vector}. Written out more explicitly, \eqref{eq:omega_omega} reads:
\begin{align}
  \omega^\nu(D_\nu\ell_\mu - D_\mu\ell_\nu - \ell_\mu D\ell_\nu) = 0\ .
\end{align}
Using these identities, the index symmetry of $\beta_{(abc)}$ and the ideal equation \eqref{eq:D_s_0}, we can expand the Chern-Simons term in \eqref{eq:div_J_1_GC} as:
\begin{align}
  \begin{split}
    & -16\mu^a\ell_\mu D_\nu(\beta_{abc}\mu^a\mu^b\omega^\nu) \\
    &\ = -\frac{64}{3} D^\nu(\beta_{abc}\mu^a\mu^b\mu^c \ell_{(\mu}\omega_{\nu)})
      + \frac{64}{3}s\ell^\nu D_{[\nu}\left(\omega_{\mu]}\frac{\beta_{abc}\mu^a\mu^b\mu^c}{s}\right). \label{eq:omega_terms}
  \end{split}
\end{align}
The first term will become part of the stress-energy conservation, while the second will be canceled by other contributions.

It remains to analyze the $\ell^\nu\del_{[\nu}c^{(1)}_{\mu]}$ term in \eqref{eq:GC_2_LHS} and the $\ell^\nu\del_{[\nu}A^{(1)a}_{\mu]}$ term in \eqref{eq:del_A_1}. They will be used to cancel the non-divergence pieces we picked up in \eqref{eq:sigma_terms} and \eqref{eq:omega_terms}. First, we gather both terms on the same side of the Gauss-Codazzi equation by moving the $\ell^\nu\del_{[\nu}A^{(1)a}_{\mu]}$ term to the LHS. We then plug in the expressions \eqref{eq:c_1} and \eqref{eq:A_1} for $c_\mu^{(1)}$ and $A^{(1)a}_\mu$, and use the ideal equation \eqref{eq:D_f} to get:
\begin{align}
  \begin{split}
    & 8\sqrt{-h}(s\ell^\nu\del_{[\nu}c^{(1)}_{\mu]} + 2\pi n_a\ell^\nu\del_{[\nu}A^{(1)a}_{\mu]}) \\
    &\ = 4\sqrt{-h}\left( \vphantom{\left(\dds{(c_a + 2\pi n^b A_{ba}/s)}\right)}
      D_\nu\left(P_\mu^\nu (s\tilde c + 2\pi n^a\tilde A_a)D_\rho\ell^\rho\right) - (\tilde c + 2\pi n^a\tilde A_a/s)D_\nu\ell^\nu D_\mu s \right. \\
    &\ \left.{} + s\left(s^2 \ddn{(c + 2\pi n^b A_b/s)}{a} - s \dds{(c_a + 2\pi n^b A_{ba}/s)} - 2\pi(\tilde A_a + s A_a)\right) D_\nu\ell^\nu D_\mu\frac{n^a}{s}\right) \\
    &\ \left.{}+ 8s\ell^\nu D_{[\nu}\left(\omega_{\mu]}(c' + 2\pi n^a A'_a/s)\right)\right. .
  \end{split}
\end{align}
This expression cancels with the non-divergence terms in \eqref{eq:sigma_terms} and \eqref{eq:omega_terms} if the coefficients in \eqref{eq:c_1} and \eqref{eq:A_1} satisfy:
\begin{align}
  & sc' + 2\pi n^a A'_a = \frac{8}{3}\beta_{abc}\mu^a\mu^b\mu^c \label{eq:cancellation_prime} \\
  & s\tilde c + 2\pi n^a\tilde A_a = -2\pi sT\mu^a\bar\sigma^{||}_{ab} \dds{(\mu^b/T)} \label{eq:cancellation_tilde} \\
  \begin{split}
    & s^2 \ddn{(c + 2\pi n^b A_b/s)}{a} - s \dds{(c_a + n^b A_{ba}/s)} - 2\pi(\tilde A_a + s A_a) \\
    &\ = 2\pi\left(\ddn{(\mu^b/T)}{c}\left(T\mu^a\bar\sigma^{||}_{ab} - s\dds{(T\mu^a\bar\sigma^{||}_{ab})}\right) + s\dds{\mu^b/T}\ddn{(T\mu^a\bar\sigma^{||}_{ab})}{c}\right). \label{eq:cancellation}
  \end{split}
\end{align}
These constraints on $(\tilde c,c,c_a,c',\tilde A_a,A_a,A_{ab},A'_a)$ clearly have non-unique solutions. Furthermore, they place no restriction on the $Z^\bot(\mathbf{n})$-part of $A_{ab}$, which affects the symmetric $Z^\bot(\mathbf{n})$-piece $\sigma^\bot_{(ab)}$ of the conductivity matrix.

Finally, we should ask whether the $\ell^\nu\del_{[\nu}c^{(1)}_{\mu]}$ and $\ell^\nu\del_{[\nu}A^{(1)a}_{\mu]}$ terms can not only cancel unwanted pieces as described above, but also generate additional contributions with the form of a symmetric tensor's divergence. If so, such contributions could be included in the hydrodynamic stress-energy conservation. The answer, however, turns out to be negative. On one hand, the $\ell^\nu\del_{[\nu}c^{(1)}_{\mu]}$ and $\ell^\nu\del_{[\nu}A^{(1)a}_{\mu]}$ terms are necessarily transverse to $\ell^\mu$. On the other hand, consider the most general first-order symmetric tensor density:
\begin{align}
  \begin{split}
    \tau^{\mu\nu} ={}& \sqrt{-h}\left(\tau_1\pi^{\mu\nu} + D_\rho\ell^\rho(\tau_2 P^{\mu\nu} + \tau_3 \ell^\mu\ell^\nu) + 2\tau_4\ell^{(\mu} D\ell^{\nu)}
	+ 2\tau^a_5\ell^{(\mu} D^{\nu)}\frac{n_a}{s}\right) \\
      & + \tau_6\ell^{(\mu}\omega^{\nu)}\ .
  \end{split}
\end{align}
The divergence of such a quantity is found, using the ideal equations, to have the following component along $\ell^\mu$:
\begin{align}
  \begin{split}
    \ell^\mu D_\nu\tau_\mu^\nu ={}& \sqrt{-h}\left((\tau_3 - \tau_1)\pi_{\mu\nu}\pi^{\mu\nu} - \tau_3\omega_{\mu\nu}\omega^{\mu\nu}
	+ \left(s\dds{\tau_3} - \frac{2}{3}\tau_3 - \tau_2\right)(D_\mu\ell^\mu)^2\right. \\
      & \left.{} -(\tau_3 + \tau_4)P^{\mu\nu}D_\mu D\ell_\nu - D\ell_\mu\left((\tau_3 + 2\tau_4) D\ell^\mu + D^\mu\tau_4\right) - \tau_3 R^{(h)}_{\mu\nu}\ell^\mu\ell^\nu \right. \\
      & \left.{} \vphantom{\left(\dds{\tau_3}\right)} - D_\mu\frac{n_a}{s}(D^\mu\tau_5^a + 2\tau_5^a D\ell^\mu) - \tau^a_5 P^{\mu\nu}D_\mu D_\nu\frac{n_a}{s}\right)
        - \omega^\mu(D_\mu\tau_6 + 3\tau_6 D\ell_\mu), \label{eq:ell_div_tau}
  \end{split}
\end{align}
where $R^{(h)}_{\mu\nu}$ is the Ricci tensor associated with the metric $h_{\mu\nu}$. From considering the independent terms in \eqref{eq:ell_div_tau}, we see that the whole expression can vanish only if all the coefficients $(\tau_1,\dots,\tau_6)$ vanish. We conclude that there is no symmetric first-order tensor whose divergence is transverse to $\ell^\mu$. Thus, one cannot generate the divergence of such a tensor from $\ell^\nu\del_{[\nu}c^{(1)}_{\mu]}$ and $\ell^\nu\del_{[\nu}A^{(1)a}_{\mu]}$ by choosing appropriate first-order corrections.

Collecting our results, we have brought the Einstein-Gauss-Codazzi equation to the form:
\begin{align}
	-8\pi D_\nu(T_\mu^{(0)\nu} + T_\mu^{(1)\nu}) = -8\pi\mu^a\ell_\mu D_\nu J_a^\nu + O(\varepsilon^3), \label{eq:div_T_1_J}
\end{align}
where the first-order stress-energy density $T_\mu^{(1)\nu}$ is given by:
\begin{align}
  \begin{split}
    T^{(1)\mu\nu} ={}& {-}\sqrt{-h}\left(\frac{s}{2\pi}\pi^{\mu\nu} + 2T\mu^b\bar\sigma^{||}_{ba}\ell^{(\mu} P^{\nu)\rho}D_\rho\frac{\mu^a}{T}\right)
	- \frac{8}{3\pi}\beta_{abc}\mu^a\mu^b\mu^c\ell^{(\mu}\omega^{\nu)} \\
      ={}& {-}\sqrt{-h}\left(\frac{s}{2\pi}\pi^{\mu\nu}
	+ \frac{s^{1/3}T}{2^{1/3}\pi}\left(\mu^a - \frac{\mu_b\mu^b n^a}{\epsilon + p}\right)\ell^{(\mu} P^{\nu)\rho}D_\rho\frac{\mu_a}{T}\right) \\
      & - \frac{8}{3\pi}\beta_{abc}\mu^a\mu^b\mu^c\ell^{(\mu}\omega^{\nu)}\ .
  \end{split}
\end{align}
Using the current conservation \eqref{eq:div_J_1_Z} and dividing by $-8\pi$, eq. \eqref{eq:div_T_1_J} becomes an energy-momentum conservation equation:
\begin{align}
	D_\nu(T_\mu^{(0)\nu} + T_\mu^{(1)\nu}) = O(\varepsilon^3). \label{eq:div_T_1}
\end{align}
The viscous stress-energy density $T_\mu^{(0)\nu} + T_\mu^{(1)\nu}$ can be written as:
\begin{align}
	T^{\mu\nu} = \sqrt{-h}\left(p h^{\mu\nu} + (\epsilon + p)u^\mu u^\nu - \frac{s}{2\pi}\pi^{\mu\nu}\right) + O(\varepsilon^2),
\end{align}
where we defined the energy velocity $u^\mu = \ell^\mu + O(\varepsilon)$ as the timelike unit eigenvector of $T^{\mu\nu}$. Explicitly, $u^\mu$ is given by:
\begin{align}
  \begin{split}
    u^\mu &= \ell^\mu - \frac{T\mu^b\bar\sigma^{||}_{ba}}{\epsilon + p} P^{\mu\nu}D_\nu\frac{\mu_a}{T}
	- \frac{4\beta_{abc}\mu^a\mu^b\mu^c}{3\pi(\epsilon + p)}\omega^\mu + O(\varepsilon^2) \\
      &= \ell^\mu - \frac{s^{1/3}T}{2^{4/3}\pi(\epsilon + p)}\left(\mu^a - \frac{\mu_b\mu^b n^a}{\epsilon + p}\right) P^{\mu\nu}D_\nu\frac{\mu_a}{T}
	- \frac{4\beta_{abc}\mu^a\mu^b\mu^c}{3\pi(\epsilon + p)}\omega^\mu + O(\varepsilon^2). \label{eq:u}
  \end{split}
\end{align}
At the given order, it makes no difference whether $\pi_{\mu\nu}$ and $\omega^\mu$ are defined in terms of $\ell^\mu$ or in terms of $u^\mu$. Using $u^\mu$ rather than $\ell^\mu$ as the basic velocity variable produces the equations in the Landau frame. We use \eqref{eq:J_1} and \eqref{eq:u} to write the conserved charge current in terms of $u^\mu$:
\begin{align}
  J_a^\mu = \sqrt{-h}\left(n_a u^\mu - T\sigma_{ab}P^{\mu\nu}D_\nu\frac{\mu^b}{T}\right)
    - \frac{2}{\pi}\left(\beta_{abc}\mu^b\mu^c - \frac{2\beta_{bcd}\mu^b\mu^c\mu^d n_a}{3(\epsilon + p)}\right)\omega^\mu + O(\varepsilon^2),
\end{align}
where the conductivity matrix is:
\begin{align}
  \sigma_{ab} &= \sigma^{||}_{ab} + \sigma^\bot_{ab}\ , \label{eq:sigma} \\
  \begin{split}
    \sigma^{||}_{ab} &\equiv \bar\sigma^{||}_{ab} - \frac{n_a\mu^c}{\epsilon + p}\bar\sigma^{||}_{cb}
      = \frac{s^{1/3}}{2^{4/3}\pi}\left(Z_{ca} - \frac{\mu_c n_a}{\epsilon + p}\right)\left(Z^c_b - \frac{\mu^c n_b}{\epsilon + p}\right) \\
      &= \frac{s^{1/3}}{2^{4/3}\pi}\left(Z_{ab} - \frac{2\mu_{(a} n_{b)}}{\epsilon + p} + \frac{\mu_c\mu^c n_a n_b}{(\epsilon + p)^2}\right). \label{eq:sigma_Z}
  \end{split}
\end{align}
The relation \eqref{eq:u} between $u^\mu$ and $\ell^\mu$ can be rewritten using $\sigma^{||}_{ab}$:
\begin{align}
  u^\mu = \ell^\mu - \frac{\mu^b\sigma^{||}_{ba}}{s} P^{\mu\nu}D_\nu\frac{\mu_a}{T}
      - \frac{4\beta_{abc}\mu^a\mu^b\mu^c}{3\pi(\epsilon + p)}\omega^\mu + O(\varepsilon^2).
\end{align}
Finally, from this we derive the expression for the entropy current in the Landau frame:
\begin{align}
  \begin{split}
    s^\mu &= \sqrt{-h}s\ell^\mu = \sqrt{-h}\left(su^\mu + \mu^b\sigma^{||}_{ba} P^{\mu\nu}D_\nu\frac{\mu_a}{T}\right)
	+ \frac{4s\beta_{abc}\mu^a\mu^b\mu^c}{3\pi(\epsilon + p)}\omega^\mu \\
      &= \sqrt{-h}\left(su^\mu + \frac{s^{4/3}T}{2^{4/3}\pi(\epsilon + p)}\left(\mu^a - \frac{\mu_b\mu^b n^a}{\epsilon + p}\right) P^{\mu\nu}D_\nu\frac{\mu_a}{T}\right)
	+ \frac{4s\beta_{abc}\mu^a\mu^b\mu^c}{3\pi(\epsilon + p)}\omega^\mu\ .
  \end{split}
\end{align}

\subsection{The focusing equation and dissipative entropy production}

It is a standard exercise to derive the second-order entropy production rate $\del_\mu s^\mu$ from the viscous hydrodynamic equations. Alternatively, it is convenient to derive entropy-related results from the focusing equation \eqref{eq:focusing}, which deals directly with the rate of area production on the horizon. We will now demonstrate this approach. For the RHS of \eqref{eq:focusing}, we have, using \eqref{eq:F_1_j_1}, \eqref{eq:mu_muT}, \eqref{eq:sigma_Z_bar}, and \eqref{eq:sigma}-\eqref{eq:sigma_Z}:
\begin{align}
  \begin{split}
    \frac{8\pi}{v}\ell^\mu F^a_{\mu\nu}j_a^\nu &= \frac{8\pi}{v^{(0)}}\ell^\mu F^{(1)a}_{\mu\nu}j_a^{(1)\nu} + O(\varepsilon^3) \\
      &= \frac{2\pi T^2}{s}\left(\frac{s^{1/3}}{2^{4/3}\pi}\left(Z_{ab} - \frac{\mu_{(a} n_{b)}}{\epsilon + p} + \frac{\mu_c\mu^c n_a n_b}{(\epsilon + p)^2}\right)
	+ \sigma^\bot_{ab}\right) P^{\mu\nu} D_\mu\frac{\mu^a}{T}D_\nu\frac{\mu^b}{T} + O(\varepsilon^3) \\
      &= \frac{2\pi T^2}{s}\sigma_{ab} P^{\mu\nu} D_\mu\frac{\mu^a}{T}D_\nu\frac{\mu^b}{T} + O(\varepsilon^3). \label{eq:bulk_T_ell_ell}
  \end{split}
\end{align}
At second order, the LHS of \eqref{eq:focusing} reads:
\begin{align}
	\kappa^{(0)}\theta - \frac{1}{(4s)^{4/3}}\sigma^{(H)}_{\mu\nu}\sigma^{(H)\mu\nu} = 2\pi T\theta - \pi_{\mu\nu}\pi^{\mu\nu} + O(\varepsilon^3),
\end{align}
where we used our choices $G^{(0)}_{\mu\nu} = h_{\mu\nu}$ and $\lambda = 1/(4s)^{2/3}$, and substituted \eqref{eq:theta_sigma_1} for $\sigma^{(H)}_{\mu\nu}$. The focusing equation then becomes:
\begin{align}
  2\pi T\theta - \pi_{\mu\nu}\pi^{\mu\nu} &= \frac{2\pi T^2}{s}\sigma_{ab} P^{\mu\nu} D_\mu\frac{\mu^a}{T}D_\nu\frac{\mu^b}{T} + O(\varepsilon^3) \\
  \Rightarrow \quad \theta &= \frac{T}{s}\sigma_{ab} P^{\mu\nu} D_\mu\frac{\mu^a}{T}D_\nu\frac{\mu^b}{T} + \frac{1}{2\pi T}\pi_{\mu\nu}\pi^{\mu\nu} + O(\varepsilon^3). \label{eq:theta2}
\end{align}
Both pieces of $\sigma_{(ab)} = \sigma^{||}_{ab} + \sigma^\bot_{(ab)}$ are positive semi-definite, as can be seen from \eqref{eq:sigma_Z} and \eqref{eq:sigma_bot_symm}. As a result, the RHS of \eqref{eq:theta2} is non-negative. The rate of area production can be derived from \eqref{eq:theta2} as:
\begin{align}
  \del_\mu S^\mu = v\theta = v^{(0)}\theta^{(2)} + O(\varepsilon^3)
    = \sqrt{-h}\left(4T\sigma_{ab} P^{\mu\nu} D_\mu\frac{\mu^a}{T}D_\nu\frac{\mu^b}{T} + \frac{2s}{\pi T}\pi_{\mu\nu}\pi^{\mu\nu}\right) + O(\varepsilon^3),
\end{align}
which translates immediately into an entropy production rate:
\begin{align}
  \del_\mu s^\mu = \frac{1}{4}\del_\mu S^\mu
    = \sqrt{-h}\left(T\sigma_{ab} P^{\mu\nu} D_\mu\frac{\mu^a}{T}D_\nu\frac{\mu^b}{T} + \frac{s}{2\pi T}\pi_{\mu\nu}\pi^{\mu\nu}\right) + O(\varepsilon^3).
\end{align}

\section{Discussion} \label{sec:discussion}

In this paper, we've analyzed the dynamics of the event horizon of a boosted Einstein-Yang-Mills black brane. The corresponding equations define the relativistic viscous hydrodynamics of conformal field theories with non-abelian conserved currents. We introduced a non-abelian Chern-Simons term and derived from the horizon dynamics the hydrodynamic constitutive relations in the presence of anomalous non-abelian global symmetries. The null nature of the event horizon plays a crucial role in the success of the derivation: the amount of independent 4d projections of the bulk fields is reduced by the relations \eqref{eq:Theta_theta} and \eqref{eq:j_F}, which are unique to null hypersurfaces.

As we have seen, the calculation lends itself to a more general context: it may be applied to the dynamics of various null hypersurfaces with Einstein-Yang-Mills fields, in an ansatz where the hypersurface's evolution equations take the form of hydrodynamic conservation laws. Equations of state other than \eqref{eq:state} can be used. Presumably, different equations of state correspond to different equilibrium bulk solutions, with eq. \eqref{eq:state} arising in the special case of the AdS black brane. Once an equation of state is chosen, in the non-abelian case there is still the freedom to choose the $Z^\bot(\mathbf{n})$-piece of the conductivity $\sigma^\bot_{ab}$. The value of $\sigma^\bot_{ab}$ arises from the corrections to the bulk gauge potential $A^a_\mu$. The standard choice in AdS/CFT arises from the condition \eqref{eq:A_infty}. Given an equilibrium bulk solution, different inhomogeneous corrections to $A^a_\mu$ lead to different non-abelian conductivities for the same equation of state (provided the corrections are such that the projected Yang-Mills equation can be cast as a conservation law).

On the other hand, we see that Einstein-Yang-Mills fields on a null horizon cannot encode an arbitrary hydrodynamic system: once the choices discussed above are made, the other transport coefficients are uniquely determined. Specifically, the shear viscosity, the bulk viscosity and the $Z(\mathbf{n})$-piece of the conductivity matrix are fixed by the null hypersurface equations to specific functions of state, while for a general fluid they may be arbitrary (under the restriction of positive semi-definiteness).

In addition, we obtained a unique form for the vorticity coefficient in the current \eqref{eq:J_1}. According to the argument presented in \cite{Son:2009tf}, this coefficient is not arbitrary for a general fluid, but uniquely determined from the anomaly coefficients of the underlying field theory. Our calculation produced the exact form for the coefficient prescribed in \cite{Son:2009tf}, with anomaly coefficients given by $C_{abc} = -(2/\pi)\beta_{abc}$. Furthermore, our result directly generalizes the result of \cite{Son:2009tf} to the case of non-abelian charges. We intend to expand on this subject in a separate work. For now, we briefly note that the argument in \cite{Son:2009tf} can be carried through for non-abelian charges by replacing all spacetime derivatives with gauge-covariant derivatives with respect to the external fields.

Our non-abelian results reduce to the abelian case in two separate ways. First, we have the case where the entire charge group is abelian to begin with. Then all charges commute with each other, and $Z^\bot(\mathbf{n})$ is the zero subspace. We therefore substitute $Z_{ab} = \delta_{ab}$ and $\sigma^\bot_{ab} = 0$ in our formulas, reproducing the known results for the abelian AdS black brane. The abelian limit should also be obtained for small charges and for weak couplings. Neglecting the $f_{abc}$-terms, eq. \eqref{eq:j_bot} becomes $j^\bot_{ab} = Z^\bot_{ab}$, while in \eqref{eq:YM_horizon_2} only $\del_\mu(j^{(1)a\mu} + \Jcs^{(1)a\mu})$ remains. Plugging in eq. \eqref{eq:j_1}, we reproduce the abelian current conservation law \eqref{eq:j_1_abelian}-\eqref{eq:div_J_1_abel}.

The hydrodynamics of the AdS black brane is conformal, due to the conformal structure of the AdS boundary. We now wish to briefly discuss the subject of conformal symmetry, which was not assumed in our derivation. For ideal hydrodynamics to be conformal, all that is required is a scale-covariant equation of state, i.e.:
\begin{align}
  \epsilon(s,n^a) = s^{4/3}\hat\epsilon(n^a/s). \label{eq:state_scaling}
\end{align}
For the (leading-order) viscous hydrodynamics to be conformal, there are two requirements. First, the shear viscosity $\eta$, the conductivity $\sigma_{ab}$ and the coefficient $\xi_a$ of the vorticity term in the charge current must likewise be scale-covariant functions:
\begin{align}
  \eta = s\hat\eta(n^a/s); \quad \sigma_{ab} = s^{1/3}\hat\sigma_{ab}(n^c/s); \quad \xi_a = s^{2/3}\hat\xi_a(n^b/s). \label{eq:transport_scaling}
\end{align}
The second requirement is for the bulk viscosity to vanish:
\begin{align}
  \zeta = 0\ . \label{eq:zeta}
\end{align}
In principle, we can have a system that is conformal in the ideal approximation, but loses conformal symmetry when the viscous corrections are taken into account. However, our results in the abelian case show that for the class of fluids described by an Einstein-Maxwell null horizon, conformal symmetry of the equation of state is sufficient for the viscous dynamics to be conformal as well. Indeed, eq. \eqref{eq:zeta} is satisfied always, while the explicit formulas for $\eta$, $\sigma_{ab}$ and $\xi_a$ guarantee that \eqref{eq:state_scaling} implies \eqref{eq:transport_scaling}. In the non-abelian case, conformal symmetry at the viscous order does not quite follow from a conformal equation of state, since $\sigma^\bot_{ab}$ can be arbitrary. This will be the case, however, if the corrections to $A^a_\mu$ which determine $\sigma^\bot_{ab}$ are governed by a conformally invariant condition, such as the condition \eqref{eq:A_infty} used in AdS/CFT.

In a previous version of this work, conformal invariance was assumed and used from the start. One can then use a Weyl-covariant formalism, as described in \cite{Loganayagam:2008is}. The main simplification in that case is that the cancellations \eqref{eq:cancellation_tilde}-\eqref{eq:cancellation} (but not \eqref{eq:cancellation_prime}) are no longer relevant: the Weyl-covariant equivalents of the corresponding terms vanish automatically, regardless of their coefficients.
		
Finally, we would like to address the effect of gauge choice on our results. The horizon fields $A^a_\mu$ and $j^a_\mu$ are subject to 4d gauge transformations, with $A^a_\mu$ transforming as a connection and $j^a_\mu$ transforming homogeneously. We are interested in descriptions where $A^a_\mu$ and $j^a_\mu$ are local functionals of the hydrodynamic fields $(h_{\mu\nu},\ell^\mu,s,n^a)$. Therefore, we are only concerned with gauge transformation that preserve this property. These are the transformations $e^{i\mathbf{\Lambda}}$ for which the angle parameter $\Lambda^a$ is itself a local functional of $(h_{\mu\nu},\ell^\mu,s,n^a)$. In general, the functional dependence of $A^a_\mu$ and $j^a_\mu$ on $(h_{\mu\nu},\ell^\mu,s,n^a)$ will be altered by such gauge transformations. The hydrodynamic equations, however, remain unchanged at the first two orders we've been considering. To see this, note that the highest-order corrections relevant for these equations are $A^{(2)a}_\mu$ and $j^{(2)a}_\mu$. Thus, we only need to consider gauge rotations with angles $\Lambda^a$ of order $1$, $\varepsilon$ and $\varepsilon^2$.

First, let us take $\Lambda^a \sim \varepsilon^2$. To second-order accuracy, the only effect of such a transformation is to rotate the zeroth-order fields $A^{(0)a}_\mu$ and $j^{(0)\mu}_a$. Now, the second-order corrections appear in our derivation only in the combination $f^a{}_{bc}(A^{(2)b}_\mu j^{(0)c\mu} + A^{(0)b}_\mu j^{(2)c\mu})$. The contribution to this quantity under the second-order gauge transformation should come from the rotation of $f^a{}_{bc}A^{(0)b}_\mu j^{(0)c\mu} = \sqrt{-h}f^a{}_{bc}\mu^b n^c$, which vanishes. We conclude that second-order gauge transformations have no effect on the hydrodynamic equations.

Let us now consider zeroth-order and first-order transformation parameters $\Lambda^a$. To maintain the first-order variable-fixing conditions of section \ref{sec:viscous:corrections}, the hydrodynamic fields must also transform under $\Lambda^a$, in the trivial ($h_{\mu\nu}$, $\ell^\mu$, $s$) and adjoint ($n^a$) representations. Now, recall that all zeroth-order and first-order functionals $\Lambda^a$ necessarily lie in $Z(\mathbf{n})$ (see section \ref{sec:viscous:class:scalars}). This means that the corresponding gauge transformations leave the fields $(h_{\mu\nu},\ell^\mu,s,n^a)$ unaltered. The hydrodynamic equations must therefore remain unchanged as well\footnote{With a little work, this can also be seen explicitly from the derivation of the equations.}.

\section*{Acknowledgements}
We would like to thank S. Minwalla for a valuable discussion.
This work is supported in part by the Israeli
Science Foundation center of excellence, by the Deutsch-Israelische
Projektkooperation (DIP), by the US-Israel Binational Science
Foundation (BSF), and by the German-Israeli Foundation (GIF).
Y.O would like to thank NORDITA and the Simons Center for Geometry and Physics for hospitality during the final stages of this work.
\appendix

\section{Notations} \label{sec:notation}

The black brane's event horizon (or a more general null hypersurface) is denoted by $\calH$. We use 5d coordinates $x^A = (r, x^\mu)$, where $r = 0$ and $\del_A r = g_{AB}\ell^B$ on $\calH$. The Yang-Mills gauge group is $\calG$.

\subsection{Index conventions}

\begin{itemize}
	\item Indices in the 5d spacetime's tensor bundle are denoted by uppercase Latin letters ($A,B,\dots$). They can be implicitly raised and lowered with the spacetime metric $g_{AB}$ and its inverse $g^{AB}$.
	\item Indices in the 4d horizon's intrinsic tensor bundle are denoted by lowercase Greek letters ($\mu,\nu,\dots$). They are implicitly raised and lowered with the ``hydrodynamic'' metric $h_{\mu\nu}$ and its inverse $h^{\mu\nu}$. Indices are never implicitly raised or lowered with the degenerate horizon metric $\gamma_{\mu\nu}$.
	\item Indices in the adjoint representation of the gauge group are denoted by lowercase Latin letters from the beginning of the alphabet ($a,b,\dots$). For non-Abelian components of the gauge group, an orthonormal basis with respect to the Killing metric is implied.
	\item Lowercase Latin indices from the middle of the alphabet $(i,j,\dots)$ run from $1$ to $\rank\calG$, and enumerate the independent scalars $N_i$ and vectors $(N_i)^a$ that can be constructed out of a gauge vector $n^a$.
\end{itemize}

\subsection{Bulk quantities}

\begin{itemize}
	\item $\epsilon^{ABCDE}$ is the 5d Levi-Civita density with components $\pm 1$. $\epsilon_{ABCDE}$ is the corresponding inverse density, also with components $\pm 1$.
	\item $g_{AB}$ is the 5d metric, with inverse $g^{AB}$ and determinant $g$. $R_{AB}$ is the corresponding Ricci tensor.
	\item $\calT_{\mu\nu}$ is the bulk stress-energy tensor.
	\item $\ell^A$ is a vector field along the horizon's null generators. $g_{AB}\ell^B$ is a covector tangent to the horizon.
\end{itemize}

\subsection{Horizon geometric quantities} \label{sec:notation:horizon}

\begin{itemize}
	\item $\epsilon^{\mu\nu\rho\sigma}$ and $\epsilon_{\mu\nu\rho\sigma}$ are the direct and inverse 4d Levi-Civita densities.
	\item $\ell^\mu$ is a vector field along the null generators.
	\item $S^\mu = v\ell^\mu$ is the area density current, independent of the scaling of $\ell^\mu$. $v$ is a scalar density that scales inversely with $\ell^\mu$.
	\item $\gamma_{\mu\nu}$ is the horizon's degenerate metric. It is the pullback of $g_{AB}$ into $\calH$.
	\item $\Theta_\mu{}^\nu = \nabla_\mu\ell^\nu$ is the Weingarten map. Its trace is $\kappa + \theta$. $\kappa$ is the surface gravity, and $\theta$ is the expansion coefficient. The lowered-index version $\Theta_\mu{}^\rho\gamma_{\rho\nu} = \theta_{\mu\nu} = \sigma^{(H)}_{\mu\nu} + (\theta/3)\gamma_{\mu\nu}$ is the horizon shear/expansion tensor. $\sigma^{(H)}_{\mu\nu}$ is the horizon shear tensor.
	\item $Q_\mu{}^\nu = v(\Theta_\mu{}^\nu - \kappa\delta_\mu^\nu)$ is the density whose divergence is used in the Gauss-Codazzi equation. $\bar Q^{\mu\nu}$ is its raised-index version with respect to $\gamma_{\mu\nu}$.
	\item $G_{\mu\nu} = \lambda\gamma_{\mu\nu} - b_\mu b_\nu$, with arbitrary scalar and covector fields $(\lambda,b_\mu)$, is an auxiliary non-degenerate metric which can replace $\gamma_{\mu\nu}$ in certain formulas. Its inverse is $(G^{-1})^{\mu\nu}$.
\end{itemize}

\subsection{Yang-Mills quantities}

\begin{itemize}
	\item $A^a_A$ is the gauge potential. $A^a_\mu$ is its pullback to $\calH$.
	\item $f_{abc} = f_{[abc]}$ are the gauge group's structure constants, with the coupling strengths included.
	\item $\beta_{abc} = \beta_{(abc)}$ are the Chern-Simons coefficients. They form a gauge-invariant constant tensor.
	\item $F^a_{AB} = 2\del_{[A}A^a_{B]} + f^a{}_{bc}A^b_A A^c_B$ is the gauge field. $F^a_{\mu\nu} = 2\del_{[\mu}A^a_{\nu]} + f^a{}_{bc}A^b_\mu A^c_\nu$ is its pullback to the horizon. $j_a^\mu = (\sqrt{-g}/4\pi)F_a^{r\mu}$ is a 4d vector density on $\calH$, related to the hydrodynamic charge current.
\end{itemize}

\subsection{Charged brane/hydrodynamic quantities}

\begin{itemize}
	\item $m$ and $q^a$ are mass and charge parameters of the black brane.
	\item $h_{\mu\nu}$ is the 4d metric of the hydrodynamics, with inverse $h^{\mu\nu}$ and determinant $h$. It can be either flat or curved.
	\item $s^\mu = \sqrt{-h}s\ell^\mu$ is the entropy current. $\ell^\mu$ is the entropy velocity, normalized with respect to $h_{\mu\nu}$. $s$ is the entropy density in its rest frame. $\ell^\mu$ is the same as the horizon's null generator defined in section \ref{sec:notation:horizon}. Note that $\ell_\mu = h_{\mu\nu}\ell^\nu$ is \emph{not} the pullback of $g_{AB}\ell^B$ to $\calH$ (which is zero). By the entropy-area relation, we have $s^\mu = S^\mu/4$.
	\item $P^\mu_\nu = \delta^\mu_\nu + \ell^\mu \ell_\nu$ is the projector orthogonal to $\ell^\mu$ (with respect to $h_{\mu\nu}$).
	\item $R$ is the brane's horizon radius, in the sense that $s = R^3/4$.
	\item $J^{a\mu}$ is the conserved hydrodynamic current. $n^a \sim q^a$ is the charge density in its rest frame. $\Jcs^{a\mu}$ is the axial part of $J^{a\mu}$, derived from the Chern-Simons term. $\sigma_{ab}$ is the conductivity matrix in the Landau frame. $\bar\sigma_{ab}$ is the conductivity matrix in the entropy frame.
	\item $T^\mu_\nu$ is the hydrodynamic stress-energy density, conserved with respect to $h_{\mu\nu}$. $u^\mu$ is the fluid's energy velocity (timelike eigenvector of $T^\mu_\nu$), normalized with respect to $h_{\mu\nu}$. $\epsilon \sim m$ is the mass/energy density in its rest frame. $p$ is the pressure.
	\item $T$ is the temperature of the brane/fluid. $\mu_a$ is the chemical potential.
	\item $Z(\mathbf{n})$ is the centralizer of $n^a$, i.e. the subspace of the gauge algebra that commutes with it. $Z_{ab}$ is the projector into $Z(\mathbf{n})$. $Z^\bot_{ab}$ is the projector into its orthogonal complement $Z^\bot(\mathbf{n})$.
	\item $N_i$ is an exhaustive set of charge scalars constructed out of $n^a$. The charge vectors $(N_i)^a = \del N_i/\del n_a$ constitute a basis for $Z(\mathbf{n})$.
\end{itemize}

\subsection{Derivatives}

\begin{itemize}
	\item $\del_\mu$ (or $\del_A$) is the ordinary partial derivative. Since we'll be dealing with distinct covariant derivatives, we use the partial-derivative symbol to highlight connection-independent objects.
	\item $\Liebold{\ell}$ is the Lie derivative with respect to $\ell^\mu$.
	\item $\nabla_A$ is the bulk covariant derivative, associated with the gravitational metric $g_{AB}$. $\bar{\nabla}_\mu$ is the restriction of $\nabla_A$ into $\calH$, in contexts where it is well-defined.
	\item $D_\mu$ is the covariant derivative associated with the 4d ``hydrodynamic'' metric $h_{\mu\nu}$. The derivative along $\ell^\mu$ is denoted by $D \equiv \ell^\mu D_\mu$.
	\item $D_\mu\ell^\mu$ is the fluid's expansion rate. $D\ell^\mu$ is the acceleration. $\pi_{\mu\nu} = P_\mu^\rho P_\nu^\sigma D_{(\rho}\ell_{\sigma)}$ is the fluid's shear tensor. $\omega_{\mu\nu} = P_\mu^\rho P_\nu^\sigma D_{[\rho}\ell_{\sigma]}$ is the vorticity. $\omega^\mu = (1/2)\epsilon^{\mu\nu\rho\sigma}\ell_\nu\omega_{\rho\sigma}$ is a vector density which also describes the vorticity. To the relevant order, these are all equivalent to the analogous quantities defined in terms of the energy velocity $u^\mu$.
	\item $D^{(G)}_\mu$ is the covariant derivative with respect to the auxiliary metric $G_{\mu\nu}$.
\end{itemize}

\end{document}